\documentclass[12pt, a4paper]{article}
\usepackage{graphicx}
\usepackage{amssymb}
\usepackage{amsmath}
\usepackage{bm}
\usepackage{color}
\usepackage{theorem}
\usepackage{subcaption}
\usepackage{listings}
\usepackage{colortbl}
\usepackage{tabularx}
\usepackage{longtable}
\usepackage[utf8]{inputenc}
\usepackage[T1]{fontenc}
\usepackage{lmodern}
\usepackage[top=30truemm,bottom=30truemm,left=25truemm,right=25truemm]{geometry}

\usepackage[sort&compress,numbers, merge]{natbib}

\definecolor{Orange}{cmyk}{0,0.61,0.87,0}
\definecolor{JungleGreen}{cmyk}{0.99,0,0.52,0}
\definecolor{OliveGreen}{cmyk}{0.64,0,0.95,0.40}
\definecolor{Brown}{cmyk}{0,0.81,1,0.60}
\definecolor{RoyalBlue}{cmyk}{0.71,0.53,0,0.12}
\definecolor{Gray}{cmyk}{0,0,0,0.40}
\definecolor{LightPink}{cmyk}{0.0,0.25,0,0}
\definecolor{LLightPink}{cmyk}{0.0,0.10,0,0}
\definecolor{LightBlue}{cmyk}{0.25,0,0,0}
\definecolor{LightGray}{cmyk}{0,0,0,0.2}

\allowdisplaybreaks[1]

\usepackage[colorlinks=true, linkcolor=OliveGreen, citecolor=RoyalBlue,
urlcolor=RoyalBlue]{hyperref}

\renewcommand{\thefootnote}{\fnsymbol{footnote}}

\begin{document}

\begin{titlepage}

  \begin{flushright}

\end{flushright}

\vskip 2.35cm
\begin{center}

{\large
{\bf
Electroweak Precision Data as a Gateway to Light Higgsinos
}
}

\vskip 1.5cm

Natsumi Nagata\footnote{
E-mail address: \href{mailto:natsumi@hep-th.phys.s.u-tokyo.ac.jp}{\tt natsumi@hep-th.phys.s.u-tokyo.ac.jp}}
and 
Genta Osaki\footnote{
  E-mail address: \href{mailto:osaki@hep-th.phys.s.u-tokyo.ac.jp}{\tt osaki@hep-th.phys.s.u-tokyo.ac.jp}}

\vskip 0.8cm

{\it Department of Physics, University of Tokyo, Bunkyo-ku, Tokyo
 113--0033, Japan} 

\date{\today}

\vskip 1.5cm

\begin{abstract}

  We investigate the prospects of probing weak-scale higgsinos through electroweak precision measurements at a future $e^+ e^-$ collider. In the Minimal Supersymmetric Standard Model, higgsinos mix with winos and binos after electroweak symmetry breaking, forming charginos and neutralinos. These states contribute to electroweak precision observables, which can be measured with high accuracy at future $e^+ e^-$ colliders. Their contributions depend on the mixing structure, as evidenced by the generation of the oblique parameters $\hat{S}$ and $\hat{T}$, in addition to the $W$ and $Y$ parameters, which arise even in the absence of mixing. We demonstrate that higgsinos with masses up to $\sim 500~\mathrm{GeV}$ can be probed through future electroweak precision experiments, highlighting their significance in probing weak-scale supersymmetry.

\end{abstract}

\end{center}
\end{titlepage}

\renewcommand{\thefootnote}{\arabic{footnote}}
\setcounter{footnote}{0}

\section{Introduction}

Higgsinos, the superpartners of the Higgs boson in supersymmetric (SUSY) models, play a crucial role in addressing the electroweak naturalness problem. Since their mass term directly contributes to the Higgs mass, it should remain close to the electroweak scale; otherwise, a large mass would require fine-tuned cancellations to achieve the correct Higgs mass~\cite{Ellis:1986yg, Barbieri:1987fn, Kitano:2005wc, Kitano:2006gv, Baer:2012up, Baer:2012cf}. This argument strongly motivates the presence of weak-scale higgsinos in SUSY extensions of the Standard Model (SM).

With $R$-parity conservation, the neutral higgsino can be a good dark matter candidate if it is the lightest SUSY particle. For higgsino masses below 1~TeV, their thermal relic abundance falls short of the observed dark matter density. This scenario, however, remains viable, as the remaining dark matter could be supplemented by non-thermal higgsino production~\cite{Gherghetta:1999sw, Moroi:1999zb, Fujii:2001xp, Gelmini:2006pw, Gelmini:2006pq, Baer:2014eja, Han:2019vxi, Fukuda:2024ddb} or consist of additional dark matter components~\cite{Bae:2013bva, Bae:2013hma, Bae:2017hlp}. 

In the Minimal SUSY SM (MSSM), higgsinos mix with winos and binos after electroweak symmetry breaking, forming charginos and neutralinos. When the gaugino masses are larger than the higgsino mass, the lightest and second-lightest neutralinos, as well as the lightest chargino, are predominantly higgsino-like, with suppressed gaugino components for larger gaugino masses. This mixing structure governs the properties of higgsino-like states and their experimental constraints~\cite{Nagata:2014wma, Fukuda:2017jmk, Krall:2017xij, Fukuda:2019kbp, Canepa:2020ntc, Martin:2024pxx, Martin:2024ytt}. For instance, limits from dark matter direct detection weaken as the gaugino masses increase and can be evaded for gaugino masses $\gtrsim 500~\mathrm{GeV}$~\cite{Martin:2024ytt}. In this regime, LHC constraints also become significantly weaker, as the small mass splittings among higgsino components ($\lesssim 10~\mathrm{GeV}$) result in soft decay products, making detection challenging. Consequently, current LHC limits on the higgsino mass in this scenario remain below 200~GeV and can be as low as 100~GeV~\cite{ATLAS:2019lng, ATLAS:2022rme, CMS:2023mny, ATLAS:2024umc}, depending on the gaugino masses.

Considering the above situation, it is crucial to thoroughly explore light higgsinos using future $e^+e^-$ colliders. Unlike dark matter direct detection experiments and LHC searches, $e^+ e^-$ collider experiments can comprehensively probe higgsino mass regions up to half of the center-of-mass energy without any coverage gaps~\cite{PardodeVera:2020zlr}. For instance, at the ILC~\cite{ILC:2013jhg, ILCInternationalDevelopmentTeam:2022izu} with $\sqrt{s} = 500~\mathrm{GeV}$, higgsinos with masses below 250 GeV can be directly searched for. A key advantage of direct searches at $e^+ e^-$ colliders is their ability to perform precise measurements of the chargino and neutralino masses, which are highly useful for investigating the underlying structure of the SUSY theory~\cite{Berggren:2013vfa, Baer:2014yta, Moortgat-Pick:2015lbx, Bae:2016dxc, Baer:2019gvu}. 

On the other hand, indirect searches for higgsinos through radiative corrections at $e^+ e^-$ colliders are also a promising avenue. Indeed, the sensitivities to new electroweakly interacting particles at $e^+ e^-$ colliders have been studied in the literature~\cite{Harigaya:2015yaa, Liu:2017msv, DiLuzio:2018jwd, Gao:2021jip}, and it has been shown that indirect searches can extend the accessible mass range beyond that of direct searches. For indirect measurements, not only high-energy linear colliders but also circular colliders with high luminosity, such as FCC-ee~\cite{FCC:2018evy, Bernardi:2022hny} and CEPC~\cite{CEPCStudyGroup:2018ghi, CEPCPhysicsStudyGroup:2022uwl}, are highly beneficial. In particular, electroweak precision measurements can play a crucial role in probing new electroweakly interacting states~\cite{Maura:2024zxz}. For recent studies on new physics searches via electroweak precision measurements at future $e^+ e^-$ colliders, see Refs.~\cite{Fan:2014vta, Fan:2014axa, Cai:2016sjz, Cai:2017wdu, Dawson:2022bxd, Bottaro:2022one, deBlas:2022ofj, Allwicher:2023aql, Bellafronte:2023amz, Allwicher:2023shc, Celada:2024mcf, Stefanek:2024kds, Knapen:2024bxw, Allwicher:2024sso, Crawford:2024nun, Erdelyi:2024sls, Greljo:2024ytg, Gargalionis:2024jaw, Davighi:2024syj, Erdelyi:2025axy}.

In the limit of heavy gaugino masses, the results of previous studies~\cite{Harigaya:2015yaa, DiLuzio:2018jwd, Maura:2024zxz} can be directly applied to higgsino searches. However, when the gaugino masses are around 1~TeV or below, the mixing between higgsinos and gauginos becomes non-negligible, influencing the properties of the higgsino-like states. In fact, the extent of this mixing significantly affects the sensitivity of electroweak precision measurements to the higgsino contribution. This can be understood by noting that in the presence of higgsino-gaugino mixing, the oblique parameters~\cite{Peskin:1991sw, Maksymyk:1993zm, Barbieri:2004qk} $\hat{S}$ and $\hat{T}$ can be induced, whereas in the limit of no mixing, only the $W$ and $Y$ parameters are generated. This distinction highlights the importance of understanding the impact of gaugino mixing when interpreting precision measurements. Motivated by this, in this paper, we investigate the sensitivity of electroweak precision measurements to higgsinos in the presence of gaugino mixing, providing insights into their discovery potential at future $e^+ e^-$ colliders. 

This paper is organized as follows. In Sec.~\ref{sec:neuandcha}, we review the neutralino and chargino sector in the MSSM and outline their gauge interactions. In Sec.~\ref{sec:vp}, we compute the contributions of neutralinos and charginos to the vacuum polarization functions of the electroweak gauge bosons. Section~\ref{sec:op} is dedicated to evaluating the oblique parameters induced by these contributions, where we compare our results with previously obtained approximated results. We then analyze the contributions to electroweak precision observables in Sec.~\ref{sec:ewpos} and demonstrate that higgsinos with masses up to $\sim 500~\mathrm{GeV}$ can be probed at future $e^+ e^-$ colliders. In Sec.~\ref{sec:high-scale_SUSY}, we discuss the impact of the modifications of the higgsino-gaugino-Higgs couplings due to a high-scale SUSY breaking on electroweak precision observables. Finally, Sec.~\ref{sec:concl} presents our conclusions. Additional useful formulae for the loop functions are provided in the Appendix.

\section{Neutralinos and charginos}
\label{sec:neuandcha}

The neutralino and chargino sectors in the MSSM consist of bino, $\widetilde{B}$, winos $\widetilde{W}^a$ ($a = 1,2,3$), and higgsinos 
\begin{equation}
    \widetilde{H}_d = 
    \begin{pmatrix}
        \widetilde{H}_d^0 \\ \widetilde{H}_d^- 
    \end{pmatrix}
    ~, \qquad 
    \widetilde{H}_u = 
    \begin{pmatrix}
        \widetilde{H}_u^+ \\ \widetilde{H}_u^0 
    \end{pmatrix}
    ~.
\end{equation}
After electroweak symmetry breaking, the neutral fields, $\widetilde{B}$, $\widetilde{W}^3$, $\widetilde{H}^0_d$, $\widetilde{H}^0_u$, and the charged fields, $\widetilde{W}^\pm$, $\widetilde{H}_d^-$, $\widetilde{H}_u^+$, mix with each other, forming neutralinos and charginos, respectively. We denote the soft SUSY-breaking mass terms of bino and winos by $M_1$ and $M_2$, respectively, and the SUSY higgsino mass term by $\mu$. 

In the weak eigenbasis $\widetilde{\psi}^0=(\widetilde{B}$, $\widetilde{W}^3$, $\widetilde{H}^0_d$, $\widetilde{H}^0_u$), the mass terms of the neutral components are collectively expressed in the form 
\begin{equation}
    \mathcal{L}_{\mathrm{mass}}^{(N)} = - \frac{1}{2} \left( \widetilde{\psi}^0 \right)^T  \mathcal{M}_N \,  \widetilde{\psi}^0 + \mathrm{h.c.} 
\end{equation}
At tree level, the neutralino mass matrix $\mathcal{M}_N$ is given by 
\begin{equation}
    \mathcal{M}_N
   =
   \begin{pmatrix}
    M_1 & 0 &  -M_Z\sin\theta_W \cos\beta   &
    M_Z\sin\theta_W\sin\beta \\
    0 & M_2 & M_Z \cos\theta_W\cos\beta  & -M_Z \cos\theta_W \sin\beta \\
    -M_Z\sin\theta_W \cos\beta & M_Z \cos\theta_W \cos\beta
    &0& -\mu \\
    M_Z \sin\theta_W \sin\beta & -M_Z \cos\theta_W \sin\beta 
   &-\mu & 0
   \end{pmatrix}
   ~,
   \label{eq:mneut}
\end{equation}
where $M_Z$ represents the mass of the $Z$ boson, $\theta_W$ is the weak mixing angle, and $\tan \beta$ denotes the ratio of the vacuum expectation values (VEVs) of the MSSM Higgs fields, $H^0_u$ and $H^0_d$: $\tan \beta \equiv \langle H^0_u\rangle / \langle H^0_d\rangle $. This matrix is diagonalized by means of a unitary
matrix $N$ as
\begin{equation}
 N^* \mathcal{M}_N N^\dagger =\text{diag}\left(M_{\widetilde{\chi}^0_1}, M_{\widetilde{\chi}^0_2}, M_{\widetilde{\chi}^0_3}, M_{\widetilde{\chi}^0_4}\right) ~,
\end{equation}
with $M_{\widetilde{\chi}^0_i} \geq 0$. The mass eigenstates, \textit{i.e.}, neutralino fields, are related to the weak eigenstates as 
\begin{equation}
    \widetilde{\chi}^0_i =N_{ij} \,\widetilde{\psi}^0_j ~.
\end{equation}

For the charged components, the mass terms are given by 
\begin{equation}
 \mathcal{L}_{\text{mass}}^{(C)}
=-
\left(\widetilde{W}^-,\widetilde{H}^-_d\right)
 \mathcal{M}_C
\begin{pmatrix}
 \widetilde{W}^+ \\ \widetilde{H}^+_u
\end{pmatrix}
+\text{h.c.}~,
\end{equation}
with
\begin{equation}
 \mathcal{M}_C=
\begin{pmatrix}
 M_2 & \sqrt{2}\,M_W \sin\beta \\
 \sqrt{2}\,M_W \cos\beta& \mu 
\end{pmatrix}
~,
\label{eq:mchar}
\end{equation}
at tree level, where $M_W$ is the $W$ boson mass. This mass  matrix is diagonalized with unitary matrices $U$ and $V$ as 
\begin{equation}
 U^*\mathcal{M}_C V^\dagger =
\text{diag}\left(M_{\widetilde{\chi}^+_1}, M_{\widetilde{\chi}^+_2}\right)
~,
\end{equation}
with the mass eigenstates (charginos) given by
\begin{equation}
 \begin{pmatrix}
  \widetilde{\chi}^+_1 \\  \widetilde{\chi}^+_2
 \end{pmatrix}
=
V
\begin{pmatrix}
 \widetilde{W}^+ \\ \widetilde{H}^+_u
\end{pmatrix}
~,~~~~~~
 \begin{pmatrix}
  \widetilde{\chi}^-_1 \\  \widetilde{\chi}^-_2
 \end{pmatrix}
=
U
\begin{pmatrix}
 \widetilde{W}^- \\ \widetilde{H}^-_d
\end{pmatrix}
~.
\end{equation}

The gauge interactions of neutralinos and charginos are 
\begin{align}
    \mathcal{L}_{\mathrm{gauge}} &= -g \,\overline{\widetilde{\chi}^+_i}
    \gamma^\mu \left(\mathcal{C}_{ij}^LP_L+\mathcal{C}_{ij}^RP_R\right)\widetilde{\chi}^0_j
    W_\mu ^++\text{h.c.} \nonumber \\ 
    &- \frac{g}{\cos \theta_W} \overline{\widetilde{\chi}^+_i}
    \gamma^\mu \left(\mathcal{G}_{ij}^LP_L+\mathcal{G}_{ij}^RP_R\right)\widetilde{\chi}^+_j
    Z_\mu 
    - \frac{g}{\cos \theta_W} \overline{\widetilde{\chi}^0_i}
    \gamma^\mu \,\mathcal{H}_{ij}^LP_L \widetilde{\chi}^0_j
    Z_\mu \nonumber \\ 
    & -e \,\overline{\widetilde{\chi}^+_i}
    \gamma^\mu \widetilde{\chi}^+_i
    A_\mu ~,
\end{align}
with 
\begin{align}
    \mathcal{C}_{ij}^L&=-V_{i1}N_{j2}^*+\frac{1}{\sqrt{2}}V_{i2}N^*_{j4}~,
   \nonumber \\ 
    \mathcal{C}_{ij}^R&=-U_{i1}^*N_{j2}-\frac{1}{\sqrt{2}}U_{i2}^*N_{j3}~, \nonumber \\ 
    \mathcal{G}^L_{ij}&=
    V_{i1}V_{j1}^* +\frac{1}{2}V_{i2}V^*_{j2}-\delta_{ij}\sin^2\theta_W~,
    \nonumber \\
     \mathcal{G}^R_{ij}&=
    U_{i1}^*U_{j1} +\frac{1}{2}U_{i2}^*U_{j2}-\delta_{ij}\sin^2\theta_W~,
    \nonumber \\
    \mathcal{H}^L_{ij}&=
    \frac{1}{2}(N_{i3}N_{j3}^*-N_{i4}N_{j4}^*)~,
\end{align}
where $g$ denotes the SU(2)$_L$ gauge coupling, $e$ is the electric charge, and we use four-component notation to describe these interaction terms.

\section{Vacuum polarization}
\label{sec:vp}

Next, we compute the contributions of charginos and neutralinos to the vacuum polarization functions of the electroweak gauge bosons, $i \Pi^{\mu\nu}_{VV'} (q)$, where $V, V' = \gamma, W, Z$. These functions are decomposed as
\begin{equation}
    \Pi^{\mu\nu}_{VV'} (q) = \Pi_{VV'} (q^2) \eta^{\mu\nu} - \Delta_{VV'} (q^2) q^\mu q^\nu ~.
\end{equation}
For electroweak precision measurements, we focus on processes where the gauge bosons couple to SM fermions that are much lighter than the electroweak scale; therefore, we can safely neglect the contribution of the second term in our calculation. We then obtain 
\begin{align}
    \Pi_{WW} (q^2) &= -\frac{g^2}{16\pi^2} \sum_{i= 1}^{2} \sum_{j=1}^{4} \Bigl[\bigl( \bigl|\mathcal{C}_{ij}^L\bigr|^2 + \bigl|\mathcal{C}_{ij}^R\bigr|^2 \bigr) H ( q^2, M^2_{\widetilde{\chi}^+_i}, M^2_{\widetilde{\chi}^0_j} ) \nonumber \\ 
    &+ 4 \mathrm{Re} \bigl( \mathcal{C}_{ij}^L \mathcal{C}_{ij}^{R*}  \bigr) M_{\widetilde{\chi}^+_i} M_{\widetilde{\chi}^0_j}B_0 ( q^2, M^2_{\widetilde{\chi}^+_i}, M^2_{\widetilde{\chi}^0_j} )
    \Bigr] ~, \\ 
    \Pi_{ZZ} (q^2) &= -\frac{g^2}{16\pi^2 \cos^2 \theta_W}  \biggl[\sum_{i,j=1}^{2} \Bigl\{
        ( \bigl|\mathcal{G}_{ij}^L\bigr|^2 + \bigl|\mathcal{G}_{ij}^R\bigr|^2 \bigr) H ( q^2, M^2_{\widetilde{\chi}^+_i}, M^2_{\widetilde{\chi}^+_j} ) \nonumber \\ 
        &+ 4 \mathrm{Re} \bigl( \mathcal{G}_{ij}^L \mathcal{G}_{ij}^{R*}  \bigr) M_{\widetilde{\chi}^+_i} M_{\widetilde{\chi}^+_j}B_0 ( q^2, M^2_{\widetilde{\chi}^+_i}, M^2_{\widetilde{\chi}^+_j} ) \Bigr\}\nonumber \\ 
        &+\sum_{i,j=1}^{4} \Bigl\{ \bigl|\mathcal{H}_{ij}^L\bigr|^2  H ( q^2, M^2_{\widetilde{\chi}^0_i}, M^2_{\widetilde{\chi}^0_j} )
        - 2\mathrm{Re} \bigl[\bigl( \mathcal{H}_{ij}^L  \bigr)^2\bigr] M_{\widetilde{\chi}^0_i} M_{\widetilde{\chi}^0_j}B_0 ( q^2, M^2_{\widetilde{\chi}^0_i}, M^2_{\widetilde{\chi}^0_j} ) \Bigr\}
    \biggr] ~, \\ 
    \Pi_{Z\gamma} (q^2) &= -\frac{eg}{16\pi^2 \cos \theta_W} \sum_{i=1}^{2} \bigl(\mathcal{G}_{ii}^L + \mathcal{G}_{ii}^R\bigr) \Bigl[H ( q^2, M^2_{\widetilde{\chi}^+_i}, M^2_{\widetilde{\chi}^+_i} ) + 2 M_{\widetilde{\chi}^+_i}^2 B_0 ( q^2, M^2_{\widetilde{\chi}^+_i}, M^2_{\widetilde{\chi}^+_i} ) \Bigr] ~, 
    \\ 
    \Pi_{\gamma\gamma} (q^2) &= -\frac{e^2}{8\pi^2 } \sum_{i=1}^{2}  \Bigl[H ( q^2, M^2_{\widetilde{\chi}^+_i}, M^2_{\widetilde{\chi}^+_i} ) + 2 M_{\widetilde{\chi}^+_i}^2 B_0 ( q^2, M^2_{\widetilde{\chi}^+_i}, M^2_{\widetilde{\chi}^+_i} ) \Bigr] ~, 
\end{align}
with
\begin{equation}
    B_0 (q^2, m_1^2, m_2^2) \equiv -\int_0^1 dx \, \ln \biggl[\frac{x\, m_1^2 + (1-x) m_2^2 - x(1-x) q^2 - i\epsilon}{Q^2}\biggr] ~,
\end{equation}
where $Q$ denotes the renormalization scale,\footnote{We take $Q = M_Z$ in the following calculation.} $\epsilon$ is a positive infinitesimal quantity, and  
\begin{align}
    H &(q^2, m_1^2, m_2^2) \equiv -\frac{1}{3} \biggl[
        m_1^2 \biggl( \frac{m_1^2-m_2^2}{q^2} - 2\biggr) \ln \biggl(\frac{m_1^2}{Q^2}\biggr)+ m_2^2 \biggl( -\frac{m_1^2-m_2^2}{q^2} - 2\biggr) \ln \biggl(\frac{m_2^2}{Q^2}\biggr) 
     \nonumber \\ 
    &-  \biggl\{
        2 q^2 - (m_1^2 + m_2^2) - \frac{(m_1^2 - m_2^2)^2}{q^2}    
        \biggr\}    B_0 (q^2, m_1^2, m_2^2)+ \frac{2q^2}{3} - \frac{(m_1^2 - m_2^2)^2}{q^2}\biggr] ~.
\end{align}
We have confirmed that these expressions are consistent with those given in Ref.~\cite{Martin:2004id} (see also Ref.~\cite{Pierce:1996zz} and references therein). We summarize some useful formulae for these loop functions in the Appendix. From Eq.~\eqref{eq:b0q0mm} and Eq.~\eqref{eq:h0q0mm} there, we find 
\begin{equation}
    \Pi_{Z\gamma} (0) = \Pi_{\gamma\gamma}(0) = 0 ~.
\end{equation}
In addition, from Eq.~\eqref{eq:b0d1q0mm} and Eq.~\eqref{eq:h0d1q0mm}, we obtain 
\begin{align}
    \Pi'_{\gamma\gamma} (0) &= \frac{e^2}{8\pi^2} \cdot \biggl(\frac{2}{3}\biggr) \sum_i \ln \biggl(\frac{M^2_{\widetilde{\chi}^+_i}}{Q^2}\biggr) ~, \\ 
    \Pi'_{Z\gamma} (0) &= \frac{e g}{16\pi^2 \cos\theta_W }  \cdot \biggl(\frac{2}{3}\biggr) \sum_{i=1}^{2} \bigl(\mathcal{G}_{ii}^L + \mathcal{G}_{ii}^R\bigr)\ln \biggl(\frac{M^2_{\widetilde{\chi}^+_i}}{Q^2}\biggr) ~.
\end{align}

\section{Oblique parameters}
\label{sec:op}

In this paper, we focus on a scenario where only neutralinos and charginos provide significant contributions to electroweak observables, assuming that all other SUSY scalar particles are sufficiently heavy. Under this assumption, the primary impact of neutralinos and charginos arises through their contributions to the vacuum polarization diagrams of electroweak gauge bosons, presented in the previous section. This effect is conveniently captured by the oblique parameters~\cite{Peskin:1991sw, Maksymyk:1993zm, Barbieri:2004qk}, defined as
\begin{align}
    \hat{S} &\equiv - \frac{\cos \theta_W}{\sin \theta_W} \Pi^{\prime}_{3B} (0) \nonumber \\ 
    &= \cos^2 \theta_W \biggl[\Pi^\prime_{ZZ} (0) - \frac{\cos^2 \theta_W - \sin^2 \theta_W}{\cos \theta_W \sin \theta_W} \Pi^{\prime}_{Z\gamma} (0) - \Pi^{\prime}_{\gamma\gamma} (0)\biggr] ~, \\ 
    \hat{T} &\equiv \frac{1}{M_W^2} \left[ \Pi_{WW} (0) - \Pi_{33} (0) \right] \nonumber \\ 
    &=  \frac{1}{M_W^2} \left[ \Pi_{WW} (0) - \cos^2 \theta_W \Pi_{ZZ} (0) \right] ~, \\ 
    W &\equiv -\frac{M_W^2}{2} \Pi^{\prime \prime}_{33} (0) \nonumber \\ 
    &= -\frac{M_W^2}{2} \left[ \cos^2 \theta_W \Pi^{\prime \prime}_{ZZ} (0) + 2 \cos \theta_W \sin \theta_W \Pi^{\prime \prime}_{Z\gamma} (0) + \sin^2 \theta_W \Pi^{\prime \prime}_{\gamma\gamma} (0) \right] ~, \\ 
    Y &\equiv -\frac{M_W^2}{2} \Pi^{\prime \prime}_{BB} (0) \nonumber \\ 
    &= -\frac{M_W^2}{2} \left[ \sin^2 \theta_W \Pi^{\prime \prime}_{ZZ} (0) - 2 \cos \theta_W \sin \theta_W \Pi^{\prime \prime}_{Z\gamma} (0) + \cos^2 \theta_W \Pi^{\prime \prime}_{\gamma\gamma} (0) \right] ~, 
\end{align}
where we follow the normalization convention in Ref.~\cite{Wells:2015uba}. These parameters are related to the Wilson coefficients of the following dimension-six effective operators 
\begin{align}
    \mathcal{L}_{\mathrm{eff}} &= C_{WB} \left( H^\dagger \sigma^a H \right) W^a_{\mu\nu} B^{\mu\nu}
    + C_T \cdot \frac{1}{2} \left( {H}^\dagger \overleftrightarrow{D}_\mu H \right)^2 \nonumber \\ 
    &- C_{2W}  \cdot \frac{1}{2} \left( D^\mu W_{\mu\nu} \right)^2 
    -  C_{2B} \cdot \frac{1}{2} \left( \partial^\mu B_{\mu\nu} \right)^2 ~,
    \label{eq:leff}
\end{align}
as 
\begin{align}
    \hat{S} = \frac{\cos \theta_W}{\sin\theta_W} v^2 C_{WB} ~, \quad 
    \hat{T} = v^2 C_T ~, \quad 
    W = M_W^2 C_{2W} ~, \quad 
    Y = M_W^2 C_{2B} ~,
\end{align}
where $H$ is the SM Higgs field, $W^a_{\mu\nu}$ and $B_{\mu\nu}$ are the field strength tensors of the $\mathrm{SU}(2)_L$ and $\mathrm{U}(1)_Y$ gauge fields, respectively, $\sigma^a$ ($a = 1,2,3$) are the Pauli matrices, ${H}^\dagger \overleftrightarrow{D}_\mu H \equiv H^\dagger D_{\mu} H - (D_\mu H)^\dagger H$, and $v \simeq 246~\mathrm{GeV}$ is the Higgs VEV. The additional electroweak oblique parameters defined in Ref.~\cite{Barbieri:2004qk}---$\hat{U}$, $V$, and $X$---are not generated at the dimension-six level and are therefore generally suppressed compared to $\hat{S}, \hat{T}, W$, and $Y$ in many new physics scenarios. Indeed, we have checked that $\hat{U}$, $V$, and $X$ remain significantly smaller than $\hat{S}, \hat{T}, W$, and $Y$ in our setup. Given their negligible impact, we omit $\hat{U}$, $V$, and $X$ from the following discussion.

Ref.~\cite{Marandella:2005wc} provides approximate expressions for the neutralino and chargino contributions to $\hat{S}, \hat{T}, W$, and $Y$, obtained by neglecting terms suppressed by $\sin^2 \theta_W$ and expanding in the ratio of the electroweak gauge boson masses to the SUSY particle masses to match the effective operators in Eq.~\eqref{eq:leff}:
\begin{align}
    \hat{S} &= \frac{g^2 M_W^2}{48\pi^2 M_2^2} \biggl[
    \frac{r (r-5-2r^2)}{(r-1)^4} + \frac{1-2r + 9r^2 -4r^3 +2r^4}{(r-1)^5} \ln r     
    \biggr] 
    \nonumber \\
    &+ \frac{g^2 M_W^2}{96\pi^2 M_2 \mu} \biggl[
        \frac{2 -19r + 20 r^2 -15 r^3}{(r-1)^4} 
        + \frac{2 + 3r -3r^2 + 4r^3}{(r-1)^5} 2r\ln r 
    \biggr]\sin 2\beta ~, \label{eq:shat_app} \\ 
    \hat{T} &= \frac{g^2 M_W^2}{192\pi^2 M_2^2} \biggl[
    \frac{7r - 29 + 16r^2}{(r-1)^3} + \frac{1 + 6r - 6r^2}{(r-1)^4} 6 \ln r    
    \biggr] \cos^2 2 \beta ~, \label{eq:that_app} \\ 
    Y &= \frac{g^{\prime 2} M_W^2}{120\pi^2 \mu^2} ~, \\
    W &= \frac{g^2}{120\pi^2} \biggl[\frac{M_W^2}{\mu^2} + \frac{2M_W^2}{M_2^2}\biggr] ~, \label{eq:w_app} 
\end{align}
where $r \equiv \mu^2/M_2^2$ and $g^\prime$ is the $\mathrm{U}(1)_Y$ gauge coupling constant. We will compare our results with these approximate formulae in the following analysis.

\begin{figure}
    \centering
    \subcaptionbox{\label{fig:shat_tanb2_mupl} $\hat{S}$ }{
    \includegraphics[width=0.47\columnwidth]{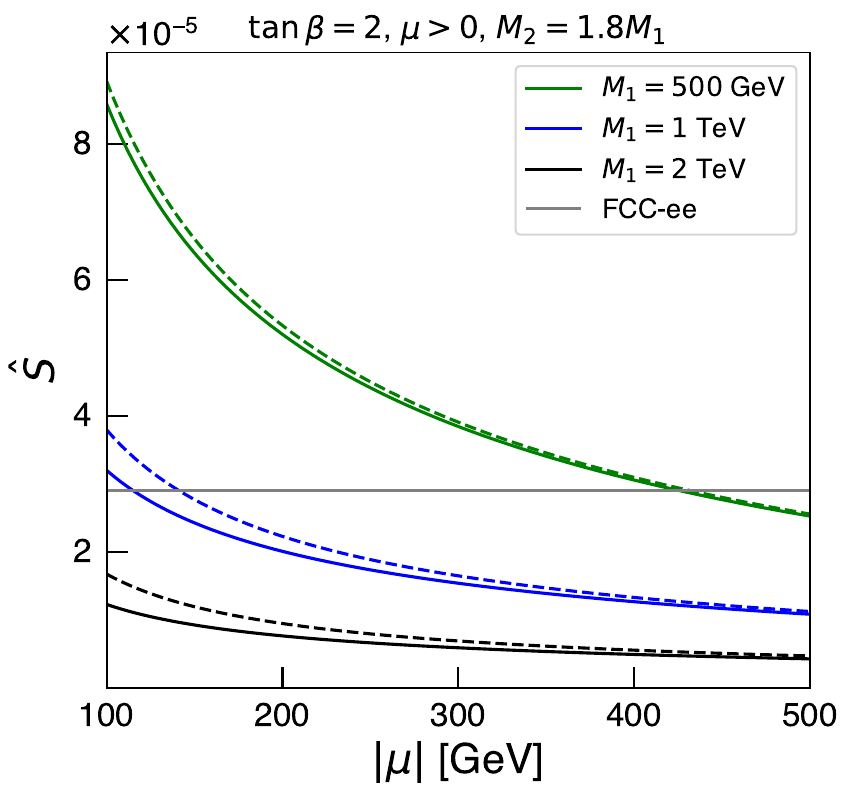}}
    \subcaptionbox{\label{fig:that_tanb2_mupl} $\hat{T}$}{
    \includegraphics[width=0.49\columnwidth]{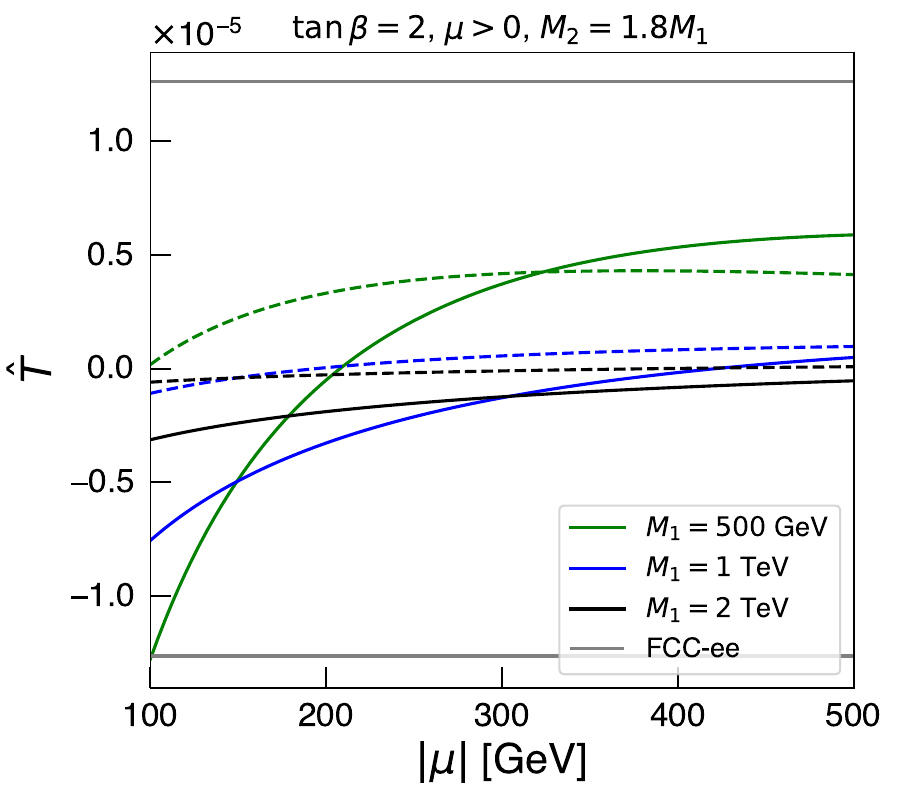}}
    \\[5pt] 
    \subcaptionbox{\label{fig:w_tanb2_mupl} $W$ }{
    \includegraphics[width=0.49\columnwidth]{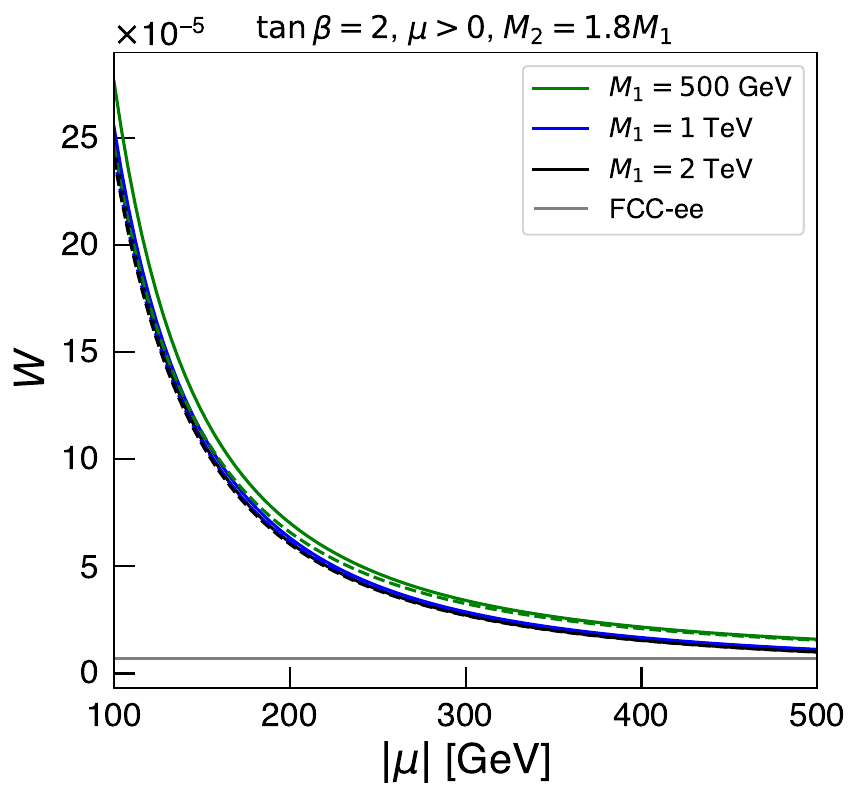}}
    \subcaptionbox{\label{fig:y_tanb2_mupl} $Y$}{
    \includegraphics[width=0.48\columnwidth]{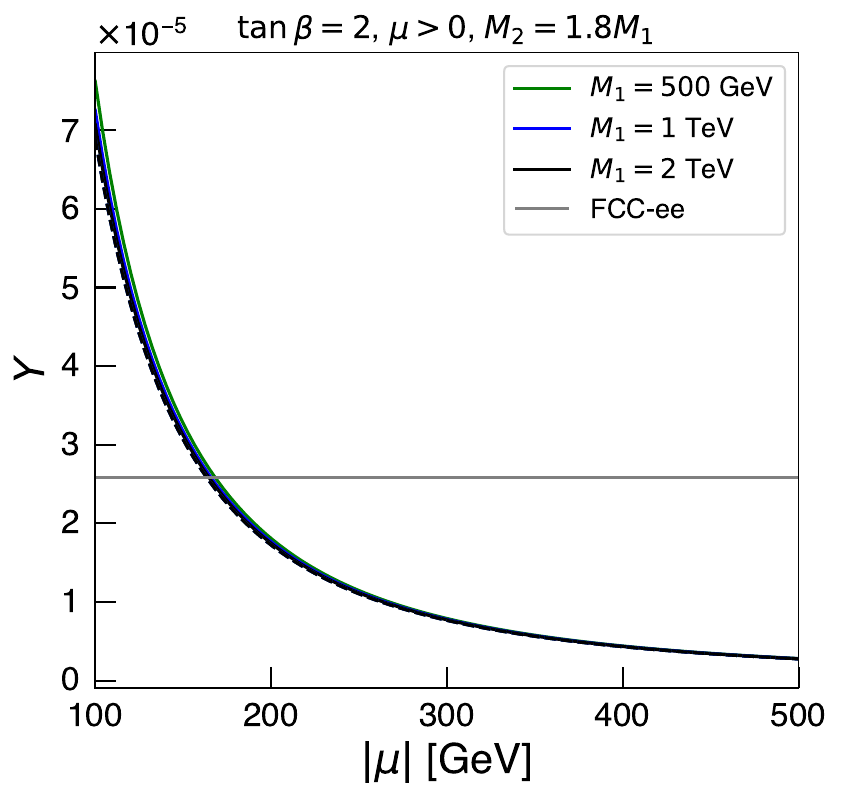}}
  \caption{
    The oblique parameters as functions of $|\mu|$ for $\tan \beta = 2$, $\mu > 0$, and $M_2 = 1.8 M_1$ in the solid curves. The blue, green, and black lines correspond to $M_1 = 500~\mathrm{GeV}$, 1~TeV, and 2~TeV, respectively. The dashed lines indicate the approximate results derived in Ref.~\cite{Marandella:2005wc}. We also depict the projected FCC-ee sensitivities, as estimated in Ref.~\cite{Maura:2024zxz}, using gray lines for reference.
  } 
\label{fig:tanb2_mupl}
\end{figure}

In Fig.~\ref{fig:tanb2_mupl}, we show the oblique parameters as functions of $|\mu|$ for $\tan \beta = 2$, $\mu > 0$, and $M_2 = 1.8 M_1$\footnote{This choice is motivated by gaugino mass unification at the GUT scale (see, \textit{e.g.}, Ref.~\cite{Martin:2024ytt}). As can be seen from Eqs.~(\ref{eq:shat_app}--\ref{eq:w_app}), our results primarily depend on the wino mass, and thus can be straightforwardly reinterpreted for different values of \(M_2/M_1\). } in the solid curves. The blue, green, and black lines correspond to $M_1 = 500~\mathrm{GeV}$, 1~TeV, and 2~TeV, respectively. The dashed lines indicate the approximate results derived in Ref.~\cite{Marandella:2005wc}. We also depict the projected FCC-ee sensitivities~\cite{Maura:2024zxz}, 
\begin{equation}
    \begin{pmatrix}
        \hat{S} \\ \hat{T} \\ W \\ Y 
    \end{pmatrix}
    = 
    \pm 
    \begin{pmatrix}
        2.907 \\ 1.264 \\ 0.681 \\ 2.583
    \end{pmatrix}
    \times 10^{-5}
    ~,
\end{equation}
using gray lines for reference. From the plots in Fig.~\ref{fig:tanb2_mupl}, we see that when the higgsino and gaugino masses are light, not only the $W$ and $Y$ parameters but also the $\hat{S}$ parameter can reach values large enough to be probed at FCC-ee. The results obtained from the approximate expressions show good agreement with our full calculations except for the $\hat{T}$ parameter. The larger relative deviation in $\hat{T}$ originates from its intrinsically small magnitude, where cancellations among the leading contributions enhance the relative impact of terms proportional to $\sin^2 \theta_W$, which are neglected in the approximate expression in Eqs.~\eqref{eq:that_app}. The $\hat{S}$ and $\hat{T}$ parameters exhibit a strong dependence on the gaugino masses, vanishing in the heavy gaugino limit. In contrast, the $W$ and $Y$ parameters show little dependence on the gaugino masses, indicating that their contributions primarily originate from the higgsino sector. Notably, the $W$ parameter suggests that FCC-ee could be sensitive to higgsino masses of up to $\sim 500~\mathrm{GeV}$, which is consistent with the result reported in Ref.~\cite{Maura:2024zxz} for an SU(2)$_L$ doublet Dirac fermion. 

\begin{figure}
    \centering
    \subcaptionbox{\label{fig:shat_tanb2_mumi} $\hat{S}$ }{
    \includegraphics[width=0.48\columnwidth]{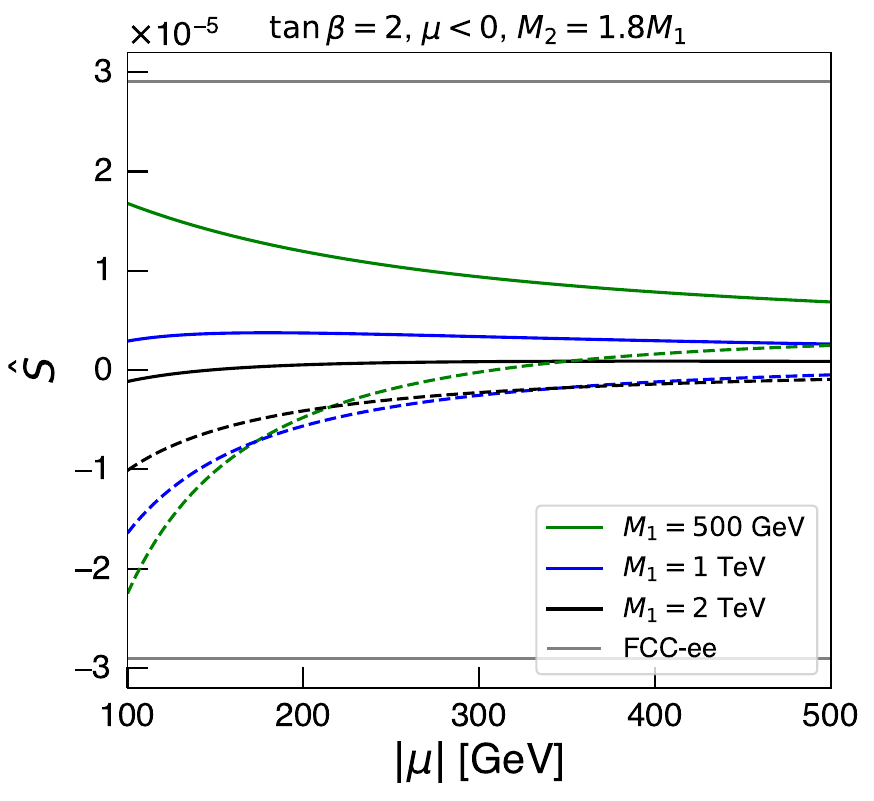}}
    \subcaptionbox{\label{fig:that_tanb2_mumi} $\hat{T}$}{
    \includegraphics[width=0.48\columnwidth]{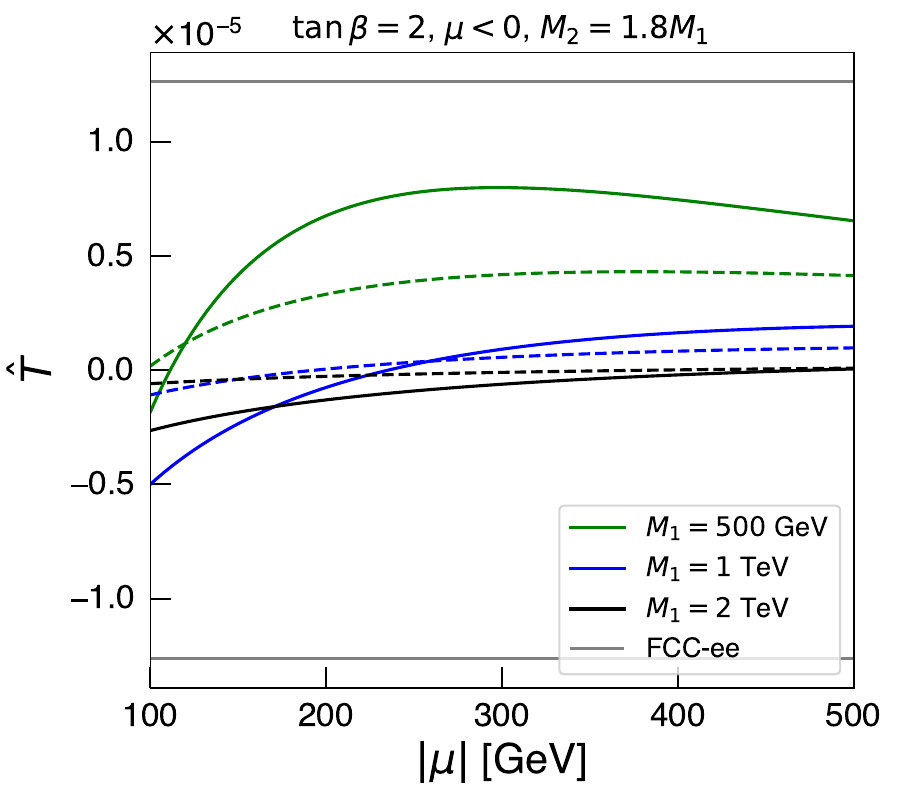}}
    \\[5pt] 
    \subcaptionbox{\label{fig:w_tanb2_mumi} $W$ }{
    \includegraphics[width=0.49\columnwidth]{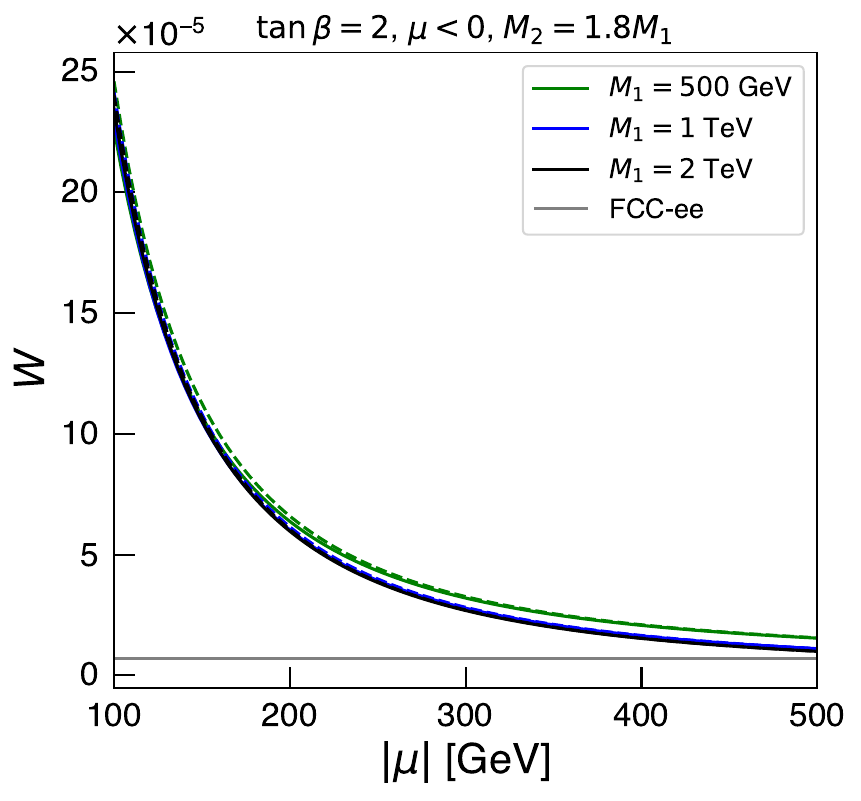}}
    \subcaptionbox{\label{fig:y_tanb2_mumi} $Y$}{
    \includegraphics[width=0.48\columnwidth]{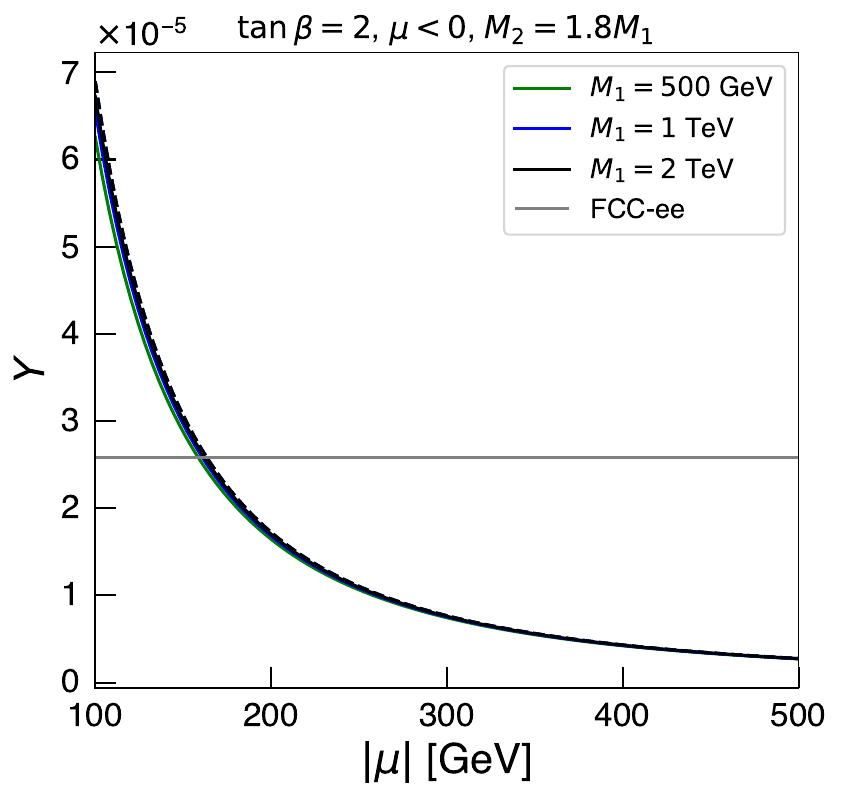}}
  \caption{
    The oblique parameters as functions of $|\mu|$ for $\tan \beta = 2$, $\mu < 0$, and $M_2 = 1.8 M_1$ in the solid curves. The blue, green, and black lines correspond to $M_1 = 500~\mathrm{GeV}$, 1~TeV, and 2~TeV, respectively. The dashed lines indicate the approximate results derived in Ref.~\cite{Marandella:2005wc}. We also depict the projected FCC-ee sensitivities, as estimated in Ref.~\cite{Maura:2024zxz}, using gray lines for reference.
  } 
\label{fig:tanb2_mumi}
\end{figure}

Figure~\ref{fig:tanb2_mumi} presents the case for $\mu < 0$, with the other parameters fixed as in Fig.~\ref{fig:tanb2_mupl}. This parameter region is known to significantly weaken constraints from dark matter direct detection experiments~\cite{Cheung:2012qy}. Compared to Fig.~\ref{fig:tanb2_mupl}, the value of the $\hat{S}$ parameter is notably smaller, while the $\hat{T}$ parameter remains small. Again, the small values arise from cancellations among the leading contributions, thereby amplifying the relative deviations between the full and approximate results. Nevertheless, even in this scenario, FCC-ee retains sensitivity to the $W$ parameter for higgsino masses $\lesssim 500~\mathrm{GeV}$.

\begin{figure}
    \centering
    \subcaptionbox{\label{fig:shat_tanb10_mupl} $\hat{S}$ }{
    \includegraphics[width=0.47\columnwidth]{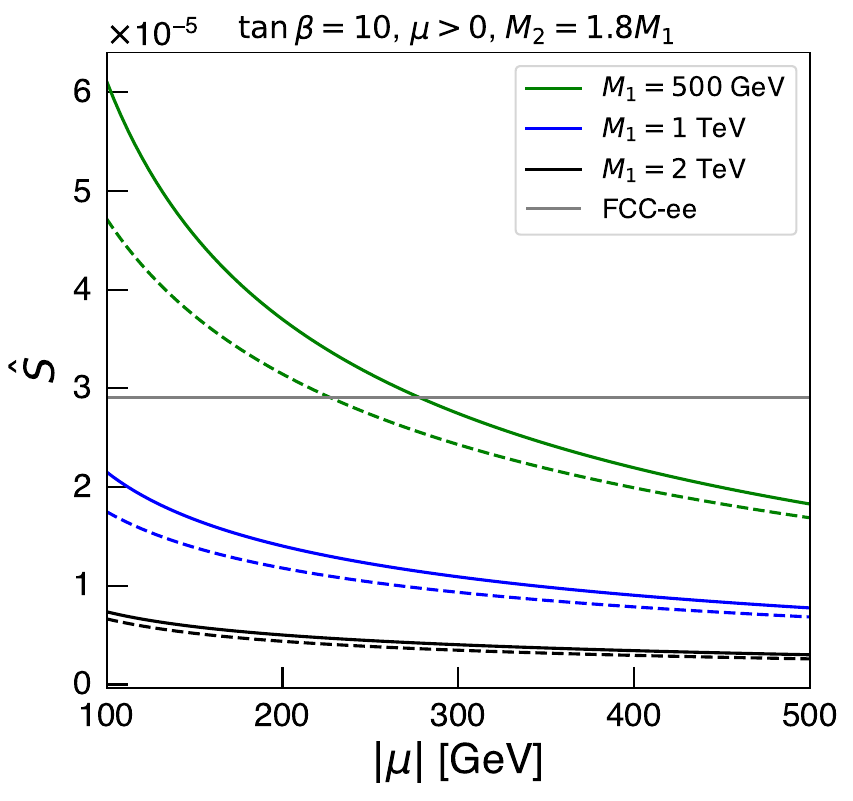}}
    \subcaptionbox{\label{fig:that_tanb10_mupl} $\hat{T}$}{
    \includegraphics[width=0.49\columnwidth]{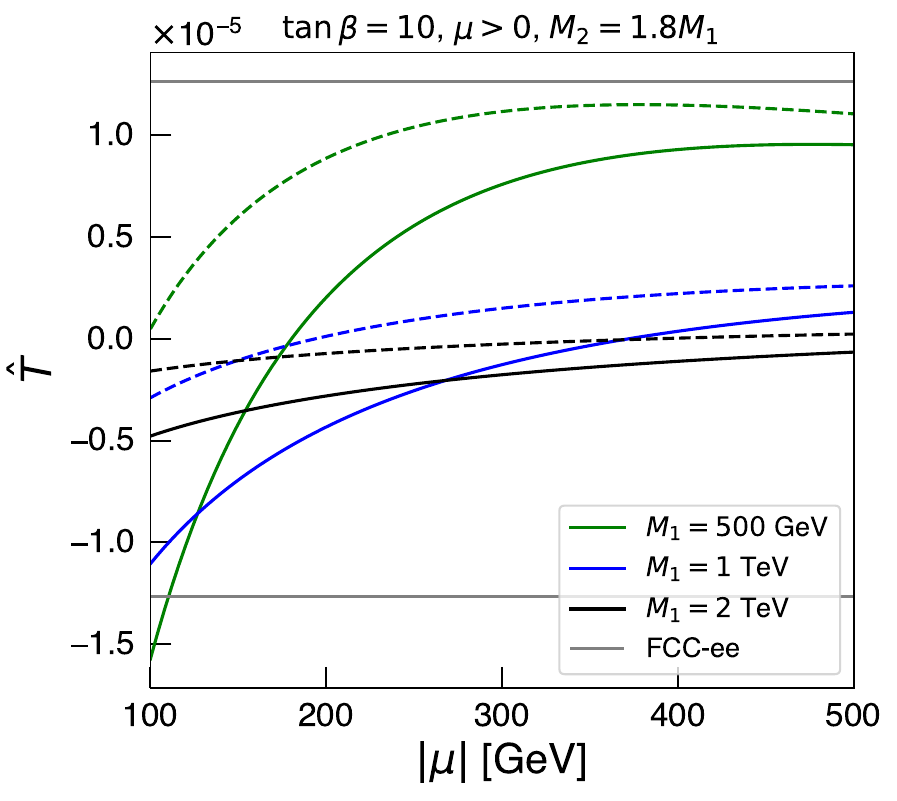}}
    \\[5pt] 
    \subcaptionbox{\label{fig:w_tanb10_mupl} $W$ }{
    \includegraphics[width=0.49\columnwidth]{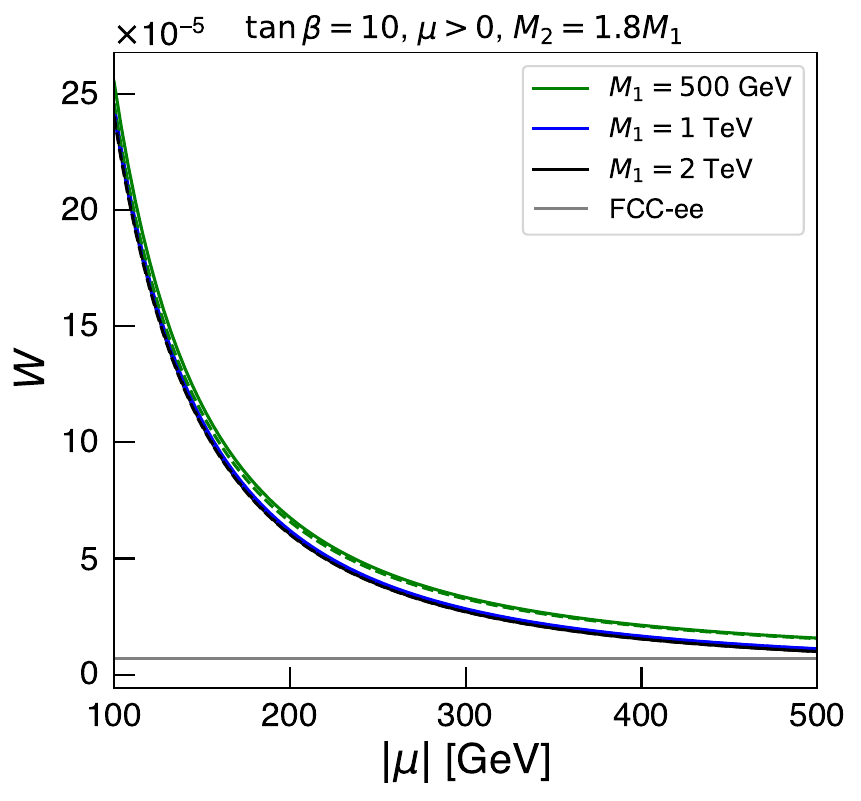}}
    \subcaptionbox{\label{fig:y_tanb10_mupl} $Y$}{
    \includegraphics[width=0.48\columnwidth]{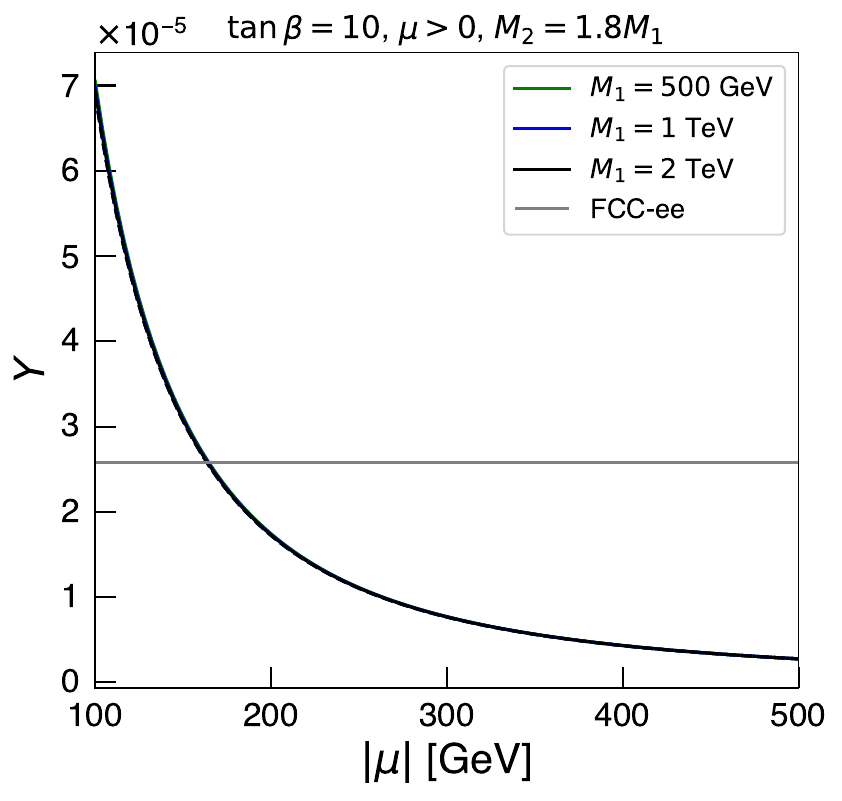}}
  \caption{
    The oblique parameters as functions of $|\mu|$ for $\tan \beta = 10$, $\mu > 0$, and $M_2 = 1.8 M_1$ in the solid curves. The blue, green, and black lines correspond to $M_1 = 500~\mathrm{GeV}$, 1~TeV, and 2~TeV, respectively. The dashed lines indicate the approximate results derived in Ref.~\cite{Marandella:2005wc}. We also depict the projected FCC-ee sensitivities, as estimated in Ref.~\cite{Maura:2024zxz}, using gray lines for reference.
  } 
\label{fig:tanb10_mupl}
\end{figure}

Figure~\ref{fig:tanb10_mupl} presents the case for \( \tan\beta = 10 \) with \( \mu > 0 \), where the parameter choices and the meaning of the curves are the same as in Fig.~\ref{fig:tanb2_mupl}. Compared to Fig.~\ref{fig:tanb2_mupl}, the value of \( \hat{S} \) is smaller, though it remains sizable when the higgsino and gaugino masses are light. In contrast, the \( W \) and \( Y \) parameters exhibit almost the same behavior as in Fig.~\ref{fig:tanb2_mupl}.

\section{Electroweak precision observables}
\label{sec:ewpos}

We now calculate the contributions of neutralinos and charginos to electroweak observables, focusing on the \( W \)-boson mass \( M_W \) and the effective weak mixing angle \( \sin^2 \theta_{\mathrm{eff}} \), representing observables from $WW$ pair production and the $Z$-pole measurements, respectively. In our analysis, we take the fine-structure constant \( \alpha \), the Fermi constant \( G_F \), and the \( Z \)-boson mass \( M_Z \), along with the SUSY parameters, as input parameters. Unless specified otherwise, the values of the input parameters are taken from from Ref.~\cite{ParticleDataGroup:2024cfk}. For previous studies on SUSY contributions to electroweak observables, see Refs.~\cite{Grifols:1984xs, Barbieri:1989dc, Drees:1990dx, Drees:1991zk, Barbieri:1991qp, Maksymyk:1993zm, Chankowski:1993eu, Pierce:1996zz, Cho:1999km, Heinemeyer:2004gx, Martin:2004id, Marandella:2005wc, Heinemeyer:2006px, Heinemeyer:2007bw}.

In terms of the oblique parameters introduced in the previous section, the shifts in \( M_W \) and \( \sin^2 \theta_{\mathrm{eff}} \) are expressed as follows: 
\begin{align}
    \frac{\Delta M_W^2}{M_W^2} &\simeq - \frac{2 \sin^2 \theta_W}{\cos 2 \theta_W} \hat{S} + \frac{\cos^2 \theta_W}{\cos 2\theta_W} \hat{T} 
    + \frac{\sin^2 \theta_W}{\cos 2 \theta_W} W + \frac{\sin^2 \theta_W}{\cos 2 \theta_W} Y ~, \label{eq:delmw}\\ 
    \frac{\Delta \sin^2 \theta_{\mathrm{eff}}}{\sin^2 \theta_{\mathrm{eff}}} &\simeq \frac{1}{\cos 2 \theta_W} \hat{S} - \frac{\cos^2 \theta_W}{\cos 2 \theta_W} \hat{T} - \frac{\sin^2 \theta_W}{\cos 2 \theta_W} W - \frac{\cos^2 \theta_W}{\cos 2\theta_W} Y ~,\label{eq:delsw}
\end{align}
where we have neglected terms proportional to the additional oblique parameters \( \hat{U} \), \( X \), and \( V \), as well as higher-order terms in the momentum expansion of the vacuum polarization functions. To obtain these relations, it is convenient to introduce the parameters $\epsilon_1, \epsilon_2, \epsilon_3$ introduced in Refs.~\cite{Altarelli:1990zd, Altarelli:1991fk}. The shifts in these parameters can be parametrized as 
\begin{align}
    \Delta \epsilon_1 &= \Delta \rho ~,  \\ 
    \Delta \epsilon_2 &= \cos^2 \theta_W \Delta \rho + \frac{\sin^2 \theta_W}{\cos 2\theta_W} \Delta r - 2 \sin^2 \theta_W \Delta \kappa ~,  \\ 
    \Delta \epsilon_3 &= \cos^2 \theta_W \Delta \rho + \cos 2 \theta_W \Delta \kappa ~, 
\end{align}
where 
\begin{equation}
    \frac{\Delta M_W^2}{M_W^2} = - \frac{\sin^2 \theta_W}{\cos 2 \theta_W} \Delta r~, \qquad 
    \frac{\Delta \sin^2 \theta_{\mathrm{eff}}}{\sin^2 \theta_{\mathrm{eff}}}  = \Delta \kappa ~.
\end{equation}
$\Delta \epsilon_{1,2,3}$ can be expressed in terms of the oblique parameters as~\cite{Barbieri:2004qk} 
\begin{align}
    \Delta\epsilon_1 &\simeq \hat{T} - W - \tan^2 \theta_W Y ~, \nonumber \\ 
    \Delta\epsilon_2 &\simeq - W  ~, \nonumber \\ 
    \Delta\epsilon_3 &\simeq \hat{S} - W - Y ~.
\end{align}
We can readily obtain Eq.~\eqref{eq:delmw} and Eq.~\eqref{eq:delsw} from these relations. 

Ref.~\cite{Martin:2004id} provides semi-analytic approximate expressions for \( \Delta M_W \) and \( \Delta \sin^2 \theta_{\mathrm{eff}} \) in the pure higgsino limit, \textit{i.e.}, in the \( M_{1,2} \to \infty \) limit:
\begin{align}
    \Delta M_W ~[\mathrm{GeV}] &= 0.00620\, r_{\widetilde{H}} + 0.00094\, r_{\widetilde{H}}^2 + 0.00017\, r_{\widetilde{H}}^3 ~,  \label{eq:delmw_app} \\ 
    \Delta \sin^2 \theta_{\mathrm{eff}} &= - 0.0000549 \, r_{\widetilde{H}} - 0.0000059 \, r_{\widetilde{H}}^2 - 0.0000009 \, r_{\widetilde{H}}^3 ~,\label{eq:delsw2_app} 
\end{align}
where $r_{\widetilde{H}} \equiv M_Z^2 /\mu^2$. We will compare our results with these approximate results in the following analysis.

\begin{table}[t]
    \centering
    \small
    \begin{tabular}{lcccccc}
    \hline \hline
    Observables & Current & HL-LHC & ILC 250 & Giga-$Z$ & CEPC & FCC-ee \\
    \hline 
      $\Delta M_W$~[MeV]  & 13.3~\cite{ParticleDataGroup:2024cfk} & 9.3~\cite{ATLAS:2018qzr} & 2.5~\cite{ILCInternationalDevelopmentTeam:2022izu} &&0.5~\cite{CEPCPhysicsStudyGroup:2022uwl}& 0.25\,(0.3)~\cite{Bernardi:2022hny} \\ 
      $\Delta \sin^2 \theta_{\mathrm{eff}}$ [$10^{-6}$] & 130~\cite{ParticleDataGroup:2024cfk} & 150~\cite{ATLAS:2018qvs, CMS:2017vxj} & 23\,(15)~\cite{Mizuno:2022xuk, ILCInternationalDevelopmentTeam:2022izu} & 4.0~\cite{ILCInternationalDevelopmentTeam:2022izu}&1.9~\cite{CEPCPhysicsStudyGroup:2022uwl}& 1.4\,(1.4)~\cite{Bernardi:2022hny} \\ 
     \hline \hline
    \end{tabular}
    \caption{The current uncertainties and the expected precision in future experiments. The values outside (inside) parentheses indicate statistical (systematic) uncertainties.
    }
    \label{tab:observables}
\end{table}

The currently measured values of $M_W$ and $\sin^2 \theta_{\mathrm{eff}}$, along with their uncertainties, are~\cite{ParticleDataGroup:2024cfk}
\begin{align}
    M_W &= 80.3692 (133)~\mathrm{GeV} ~, \nonumber \\ 
    \sin^2 \theta_{\mathrm{eff}} &= 0.23149(13) ~.
\end{align}
We summarize the current uncertainties of these observables and the expected precision in future experiments in Table~\ref{tab:observables}. In some cases, only the expected precision of the electron left-right symmetry $A_e$ is provided, and not that of $\sin^2 \theta_{\mathrm{eff}}$. In such cases, the precision of $\sin^2 \theta_{\mathrm{eff}}$ is estimated using the relation
\begin{equation}
    A_e = \frac{1 - 4 \sin^2 \theta_{\mathrm{eff}}}{1 -4 \sin^2 \theta_{\mathrm{eff}} + 8\sin^4 \theta_{\mathrm{eff}}} ~,
\end{equation}
namely, 
\begin{align}
    \Delta A_e = \frac{-8 + 8(1-4\sin^2 \theta_{\mathrm{eff}})^2}{[1 + (1- 4\sin^2 \theta_{\mathrm{eff}})^2]^2} \Delta \sin^2 \theta_{\mathrm{eff}}
    \simeq - 7.9 \Delta \sin^2 \theta_{\mathrm{eff}} ~.
    \label{eq:delaedels2}
\end{align}

For the $W$-boson mass measurement at the HL-LHC, we quote the precision estimated for the lepton pseudorapidity coverage of $|\eta_\ell| < 4$ and a data set of 200~pb$^{-1}$~\cite{ATLAS:2018qzr}. Regarding $\Delta \sin^2 \theta_{\mathrm{eff}}$ at the HL-LHC, we show the ATLAS projections~\cite{ATLAS:2018qvs} based on the PDF4LHC15$_{\text{HL-LHC}}$ set~\cite{AbdulKhalek:2018rok}; similar sensitivities are expected from CMS, with $\Delta \sin^2 \theta_{\mathrm{eff}} = (12-15) \times 10^{-5}$~\cite{CMS:2017vxj}. 

At the ILC with \( \sqrt{s} = 250~\mathrm{GeV} \), the measurement of the \( W \)-boson mass is primarily limited by systematic uncertainties. However, by combining multiple measurement methods, it is expected that \( M_W \) can be determined with a precision of at least 2.5~MeV~\cite{ILCInternationalDevelopmentTeam:2022izu}. A dedicated run near the \( W^+ W^- \) production threshold can also be utilized to measure the \( W \)-boson mass. With \( 500~\mathrm{fb}^{-1} \) of data, an experimental precision of $\simeq 2~\mathrm{MeV}$ is expected to be achievable~\cite{Wilson:2016hne}. On the other hand, \( \sin^2 \theta_{\mathrm{eff}} \) can be measured through the radiative return process, \( e^+ e^- \to Z \gamma \). The expected statistical and systematic uncertainties in determining \( A_e \) using this method are \( 1.8 \times 10^{-4} \)~\cite{Mizuno:2022xuk} and \( 1.2 \times 10^{-4} \)~\cite{ILCInternationalDevelopmentTeam:2022izu}, respectively. We use the relation~\eqref{eq:delaedels2} to estimate $\Delta \sin^2 \theta_{\mathrm{eff}}$ in Table~\ref{tab:observables}. 

For \( \sin^2 \theta_{\mathrm{eff}} \), the Giga-Z program---an operation of the ILC at the \( Z \) pole~\cite{Yokoya:2019rhx}---offers an approximately one-order-of-magnitude improvement in sensitivity. With \( 100~\mathrm{fb}^{-1} \) data-taking for each helicity configuration and standard beam polarizations, the combined statistical and systematic uncertainty on \( \sin^2 \theta_{\mathrm{eff}} \) is estimated to be \( \Delta \sin^2 \theta_{\mathrm{eff}} = 4.0 \times 10^{-6} \)~\cite{ILCInternationalDevelopmentTeam:2022izu}. We also note that in the most optimistic scenario considered in Ref.~\cite{ILCInternationalDevelopmentTeam:2022izu}, $A_e$ can be measured with both the statistical and systematic uncertainties being $\Delta A_e = 1.3 \times 10^{-5}$, which corresponds to \( \Delta \sin^2 \theta_{\mathrm{eff}} = 1.7 \times 10^{-6} \).

Circular colliders offer superior precision in measuring both \( M_W \) and \( \sin^2 \theta_{\mathrm{eff}} \). At CEPC, expected uncertainties of \( \Delta M_W = 0.5~\mathrm{MeV} \) and \( \Delta A_e = 1.5 \times 10^{-5} \) can be achieved~\cite{CEPCPhysicsStudyGroup:2022uwl}; using Eq.~\eqref{eq:delaedels2}, the latter corresponds to \( \Delta \sin^2 \theta_{\mathrm{eff}} = 1.9 \times 10^{-6} \). The projected sensitivities of FCC-ee for \( M_W \) and \( \sin^2 \theta_{\mathrm{eff}} \) in Table~\ref{tab:observables} are taken from Ref.~\cite{Bernardi:2022hny}.

\begin{figure}
    \centering
    \subcaptionbox{\label{fig:delmw_tanb2_mupl} $\mu > 0$ }{
    \includegraphics[width=0.48\columnwidth]{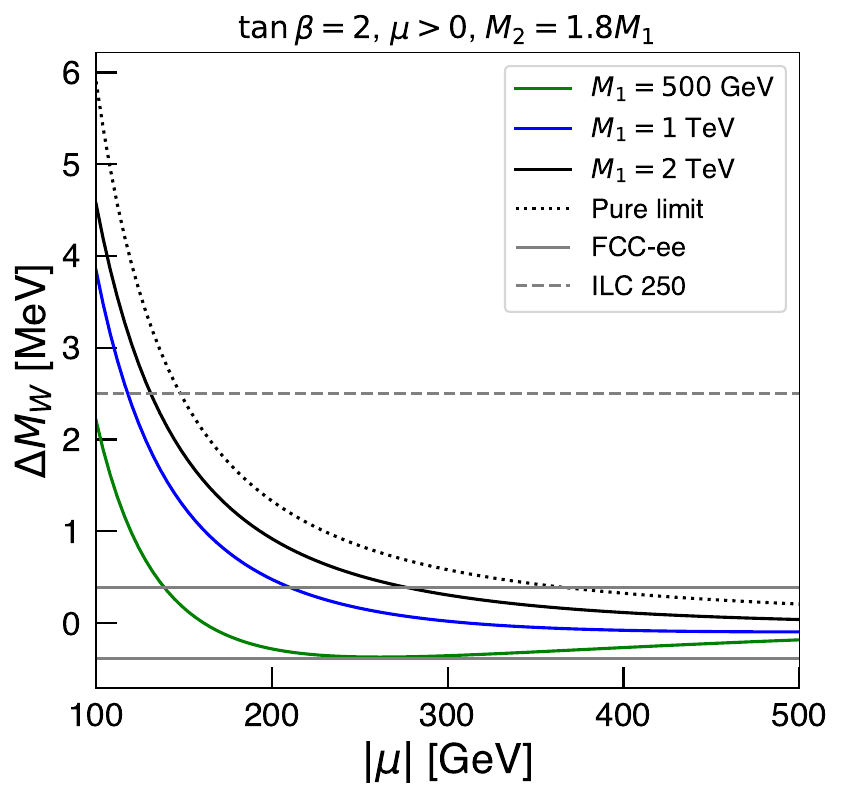}}
    \subcaptionbox{\label{fig:delmw_tanb2_mumi} $\mu < 0$}{
    \includegraphics[width=0.48\columnwidth]{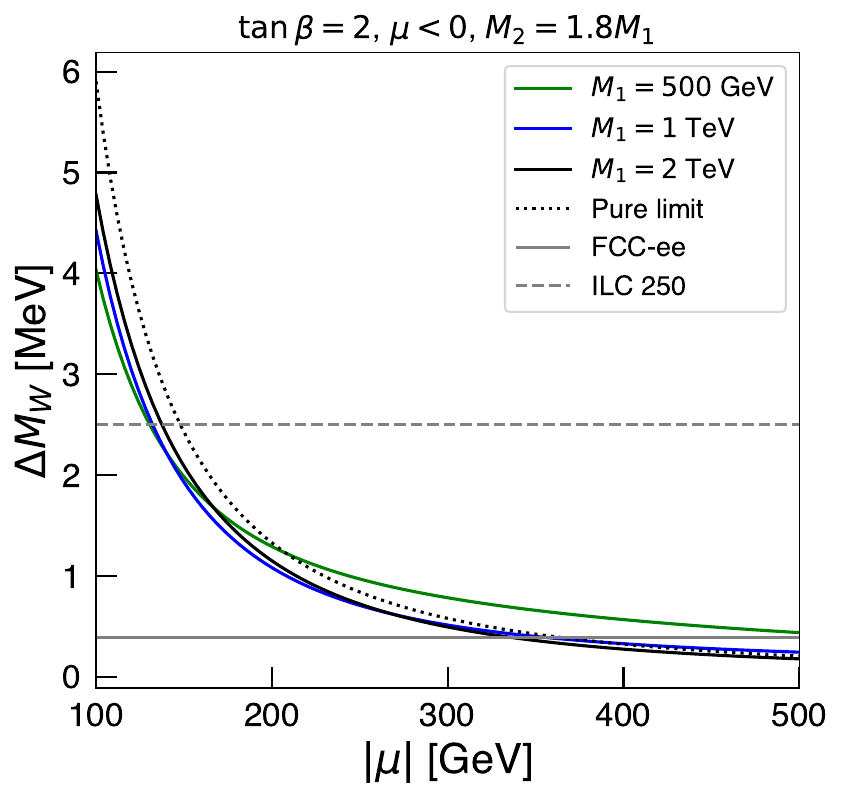}}
  \caption{\( \Delta M_W \) as a function of \( |\mu| \), with \( \tan \beta = 2 \) and \( M_2 = 1.8 M_1 \). The left (right) panel corresponds to the case of \( \mu > 0 \) (\( \mu < 0 \)). The green, blue, and black solid lines represent \( M_1 = 500~\mathrm{GeV} \), 1~TeV, and 2~TeV, respectively. The black dotted line indicates the pure higgsino limit given in Eq.~\eqref{eq:delmw_app}. The horizontal gray solid (dashed) lines denote the projected sensitivity of FCC-ee (ILC 250) as summarized in Table~\ref{tab:observables}.
  } 
\label{fig:delmw}
\end{figure}

In Fig.~\ref{fig:delmw}, we present \( \Delta M_W \) as a function of \( |\mu| \), with \( \tan \beta = 2 \) and \( M_2 = 1.8 M_1 \). The left (right) panel corresponds to the case of \( \mu > 0 \) (\( \mu < 0 \)). The green, blue, and black solid lines represent \( M_1 = 500~\mathrm{GeV} \), 1~TeV, and 2~TeV, respectively. The black dotted line indicates the pure higgsino limit given in Eq.~\eqref{eq:delmw_app}. The horizontal gray solid (dashed) lines denote the projected sensitivity of FCC-ee (ILC 250) as summarized in Table~\ref{tab:observables}. For the FCC-ee sensitivity, we combine the statistical and systematic uncertainties in quadrature. Although omitted from the figure for clarity, the sensitivity of CEPC is comparable to that of FCC-ee, as can be seen in Table~\ref{tab:observables}. Figure~\ref{fig:delmw_tanb2_mupl} shows that for \( \mu > 0\) and the gaugino masses \( \simeq 1~\mathrm{TeV}\), \( \Delta M_W \) is reduced compared to the pure higgsino limit. Consequently, while FCC-ee could probe higgsino masses up to \(\simeq 350~\mathrm{GeV}\) in the pure higgsino limit, the sensitivity decreases when the gaugino mass is \(\lesssim 1~\mathrm{TeV}\), restricting the search to lighter higgsino masses. We also find that, depending on the gaugino mass, ILC 250 may be able to probe higher higgsino masses via electroweak precision measurements than through direct production, which is limited to $< 125$~GeV. On the other hand, as shown in Fig.~\ref{fig:delmw_tanb2_mumi}, for \( \mu < 0 \), the dependence of \( \Delta M_W \) on the gaugino masses is significantly reduced. Notably, the FCC-ee sensitivity to higgsino masses can be comparable to or even exceed that of the pure higgsino limit (\(\simeq 350\)~GeV).

\begin{figure}
    \centering
    \subcaptionbox{\label{fig:delsw2_tanb2_mupl} $\mu > 0$ }{
    \includegraphics[width=0.48\columnwidth]{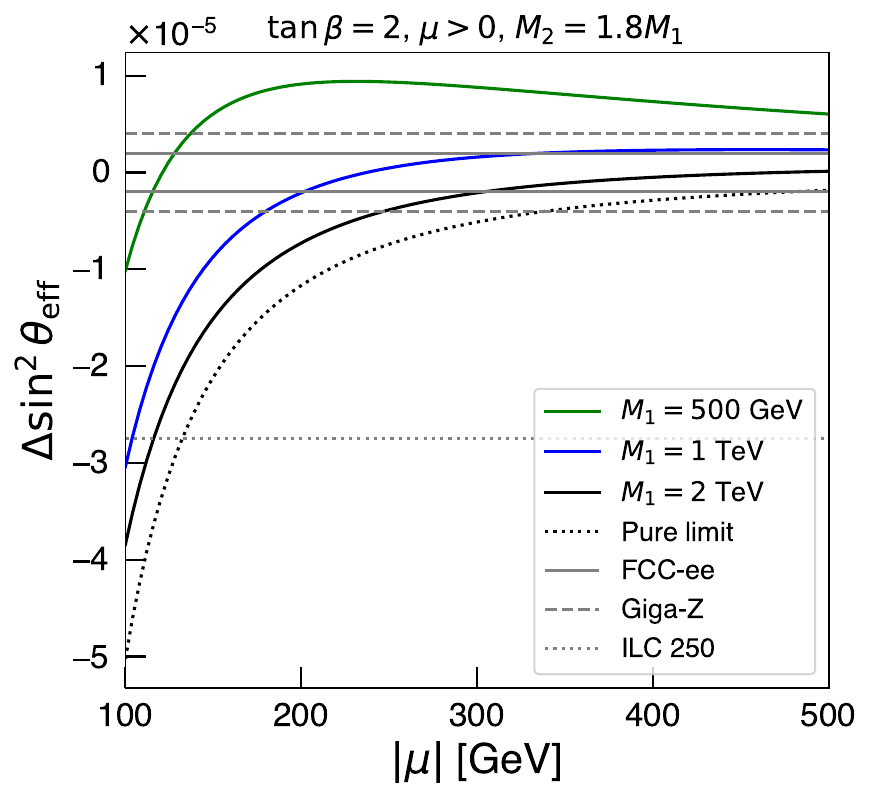}}
    \subcaptionbox{\label{fig:delsw2_tanb2_mumi} $\mu < 0$}{
    \includegraphics[width=0.48\columnwidth]{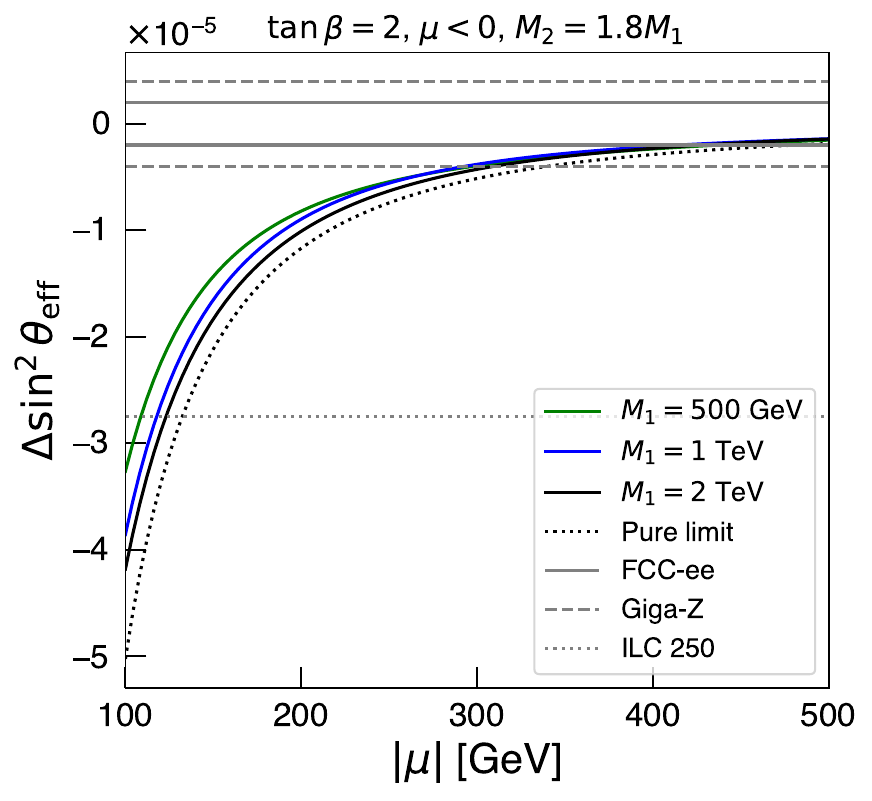}}
  \caption{
    \( \Delta \sin^2 \theta_{\mathrm{eff}} \) as a function of \( |\mu| \), with \( \tan \beta = 2 \) and \( M_2 = 1.8 M_1 \). The left (right) panel corresponds to the case of \( \mu > 0 \) (\( \mu < 0 \)). The green, blue, and black solid lines represent \( M_1 = 500~\mathrm{GeV} \), 1~TeV, and 2~TeV, respectively, while the black dotted line indicates the pure higgsino limit given in Eq.~\eqref{eq:delsw2_app}. The horizontal gray solid, dashed, and dotted lines represent the projected sensitivities of FCC-ee, Giga-Z, and ILC 250, respectively, as summarized in Table~\ref{tab:observables}.
  } 
\label{fig:delsw2}
\end{figure}

In Fig.~\ref{fig:delsw2}, we show \( \Delta \sin^2 \theta_{\mathrm{eff}} \) as a function of \( |\mu| \), with \( \tan \beta = 2 \) and \( M_2 = 1.8 M_1 \). The left (right) panel corresponds to the case of \( \mu > 0 \) (\( \mu < 0 \)). The green, blue, and black solid lines represent \( M_1 = 500~\mathrm{GeV} \), 1~TeV, and 2~TeV, respectively, while the black dotted line indicates the pure higgsino limit given in Eq.~\eqref{eq:delsw2_app}. The horizontal gray solid, dashed, and dotted lines represent the projected sensitivities of FCC-ee, Giga-Z, and ILC 250, respectively, as summarized in Table~\ref{tab:observables}. Again, for the FCC-ee sensitivity, we combine the statistical and systematic uncertainties in quadrature, and we omit the CEPC sensitivity from the plots for clarity, which is comparable to that of FCC-ee.  Similar to the case of \( \Delta M_W \), the dependence on the gaugino mass is significant for \( \mu > 0 \), whereas it becomes much weaker for \( \mu < 0 \). In the pure higgsino limit, FCC-ee is expected to probe higgsino masses up to $\simeq 500~\mathrm{GeV}$. However, for gaugino masses around 1 TeV, the sensitivity of FCC-ee may be reduced for \( \mu > 0 \).  Moreover, Giga-Z can probe higgsino masses in the several hundred GeV range for a pure higgsino, covering a broader parameter space than direct higgsino searches, not only at ILC 250 but even at ILC 500.

\section{High-scale SUSY}
\label{sec:high-scale_SUSY}

In the above analysis, we have assumed that all SUSY scalar particles are sufficiently heavy, rendering their contributions to electroweak precision observables negligible. A well-motivated scenario for this mass hierarchy is one in which fermionic SUSY particles are lighter than the SUSY scalar particles by a one-loop factor, as considered in Refs.~\cite{Wells:2003tf, Wells:2004di, Arkani-Hamed:2004ymt, Giudice:2004tc, Arkani-Hamed:2004zhs, Arkani-Hamed:2005zuc, Hall:2011jd, Hall:2012zp, Ibe:2011aa, Ibe:2012hu, Arvanitaki:2012ps, Arkani-Hamed:2012fhg, Evans:2013lpa, Evans:2013dza}. In this setup, if the fermionic SUSY particles have masses around 1~TeV, the SUSY scalar particles are expected to lie in the range of \( \mathcal{O}(100\text{--}1000)~\mathrm{TeV} \).  

Under these conditions, the higgsino-gaugino-Higgs couplings, which are determined by gauge couplings in SUSY theories, generally deviate from their SUSY-predicted relations at low energies~\cite{Arkani-Hamed:2004ymt, Giudice:2004tc}. This deviation affects the off-diagonal components of the neutralino and chargino mass matrices, \( \mathcal{M}_N \) and \( \mathcal{M}_C \), where the expressions in Eq.~\eqref{eq:mneut} and Eq.~\eqref{eq:mchar} are derived under the assumption that the SUSY relations hold. The higgsino-gaugino-Higgs interactions below the SUSY scale are described by 
\begin{align}
  \mathcal{L}_{\mathrm{int}} = &- \frac{\tilde{g}_u }{\sqrt{2}} H^\dagger \sigma^a \widetilde{H}_u \widetilde{W}^a  - \frac{\tilde{g}^\prime_u }{\sqrt{2}} H^\dagger  \widetilde{H}_u \widetilde{B} \nonumber \\ 
  &+ \frac{\tilde{g}_d }{\sqrt{2}} \epsilon^{\alpha \beta} (H)_\alpha (\sigma^a \widetilde{H}_d)_\beta \widetilde{W}^a- \frac{\tilde{g}^\prime_d }{\sqrt{2}} \epsilon^{\alpha \beta} (H)_\alpha(\widetilde{H}_d)_\beta\widetilde{B}  + \mathrm{h.c.}
\end{align}
At the SUSY scale, the coefficients of these terms are given as 
\begin{align}
  \tilde{g}_u &= g \sin \beta ~, \qquad \tilde{g}'_u = g' \sin \beta ~, \qquad 
  \tilde{g}_d = g \cos \beta ~, \qquad \tilde{g}'_d = g' \cos \beta ~. 
\end{align}
These couplings are evolved down to the gaugino mass scale using the renormalization group equations (RGEs). As shown in Ref.~\cite{Giudice:2004tc}, for the SUSY scales $\lesssim 1000~\mathrm{TeV}$, $\tilde{g}_u \simeq g \sin \beta$ and $\tilde{g}_d \simeq g \cos \beta$, but $\tilde{g}'_u /(g' \sin \beta)$ and $\tilde{g}'_d /(g' \cos \beta)$ can decrease by $\sim 10$\%. 

The impact of these coupling modifications on electroweak precision observables is most pronounced when \( \hat{S} \) or \( \hat{T} \) takes large values, as the effect arises through higgsino-gaugino mixing. To assess this, therefore, we focus on the case where gauginos are light, with \( \mu > 0 \) and \( \tan \beta = 2 \). We have found that in this case, the shifts in \( \Delta M_W \) and \( \Delta \sin^2 \theta_{\mathrm{eff}} \) from the results in the previous section are at most a few percent, which is significantly smaller than the experimental sensitivities. This indicates that the RGE effects on these couplings can be safely neglected in our analysis.

\section{Conclusion}
\label{sec:concl}

In this work, we have investigated the potential of electroweak precision measurements at future \( e^+ e^- \) colliders to probe light higgsinos in the MSSM. Unlike direct searches at hadron colliders, which face challenges due to the compressed mass spectrum of higgsinos, electroweak precision observables provide an indirect yet powerful avenue to detect their effects through loop-induced contributions.

We have studied the contributions of neutralinos and charginos to the oblique parameters \( \hat{S} \), \( \hat{T} \), \( W \), and \( Y \). Our analysis shows that \( \hat{S} \) and \( \hat{T} \) are highly sensitive to the gaugino mass scale and can become comparable in magnitude to \( W \) and \( Y \) when gaugino masses are around the TeV scale. By examining the impact of these contributions on the precision observables \( M_W \) and \( \sin^2 \theta_{\mathrm{eff}} \), we have demonstrated that future \(e^+e^-\) colliders such as the ILC, FCC-ee, and CEPC possess significant sensitivity to light higgsinos, with the potential to probe masses up to \(\simeq 500~\mathrm{GeV}\). Furthermore, we have shown that the Giga-Z program at the ILC, through its high-precision measurement of \( \sin^2 \theta_{\mathrm{eff}} \), can probe higgsino masses in the several hundred GeV range, potentially exceeding the reach of direct higgsino searches at ILC 500.

\section*{Acknowledgments}

The work of NN was supported in part by the Grant-in-Aid for Young Scientists (No.~21K13916).  

\section*{Appendix: formulae for the loop functions}
\renewcommand{\theequation}{A.\arabic{equation}}
\setcounter{equation}{0}

We summarize some useful formulae for the loop functions and their derivatives. For $q^2 = 0$, we have   
\begin{align}
    B_0 (0, m_1^2, m_2^2) &= 1- \frac{1}{m_1^2 - m_2^2} \biggl[m_1^2 \ln \biggl(\frac{m_1^2}{Q^2} \biggr) - m_2^2 \ln \biggl(\frac{m_2^2}{Q^2}\biggr) \biggr] ~, \\ 
    B_0 (0, m^2, m^2) &= - \ln \biggl(\frac{m^2}{Q^2}\biggr) ~, \label{eq:b0q0mm} \\ 
    H (0, m_1^2, m_2^2) &=- \frac{m_1^2 + m_2^2}{3} \nonumber \\ &+ \frac{1}{3(m_1^2 - m_2^2)} \biggl[m_1^2 \left( 3 m_1^2 - m_2^2 \right) \log \biggl(\frac{m_1^2}{Q^2}\biggr) + m_2^2 \left( m_1^2 -3 m_2^2 \right) \log \biggl(\frac{m_2^2}{Q^2}\biggr) \biggr] ~, \\ 
    H (0, m^2, m^2) &=2 m^2 \log \biggl(\frac{m^2}{Q^2}\biggr) ~. \label{eq:h0q0mm}
\end{align}

The first derivatives of the functions $B_0 (q^2, m_1^2, m_2^2)$ and $H_0 (q^2, m_1^2, m_2^2)$ with respect to $q^2$ are given by 
\begin{align}
    B_0' (q^2, m_1^2, m_2^2) &\equiv
    \frac{d}{dq^2} B_0 (q^2, m_1^2, m_2^2) = \int_0^1 dx \, \frac{x(1-x) }{x m_1^2 + (1-x) m_2^2 - x (1-x) q^2 - i\epsilon} ~, \\
    H' (q^2, m_1^2, m_2^2) &\equiv
    \frac{d}{dq^2} H_0 (q^2, m_1^2, m_2^2) \nonumber \\ 
    &= - \frac{1}{3} \biggl[
    - \frac{m_1^2 (m_1^2 - m_2^2)}{(q^2)^2} \ln \biggl(\frac{m_1^2}{Q^2}\biggr)  + \frac{m_2^2 (m_1^2 - m_2^2)}{(q^2)^2} \ln \biggl(\frac{m_2^2}{Q^2}\biggr) +\frac{2}{3}  \nonumber \\ &+ \frac{(m_1^2-m_2^2)^2}{(q^2)^2}
    - \biggl\{2+\frac{(m_1^2-m_2^2)^2}{(q^2)^2}\biggr\} B_0(q^2,m_1^2,m_2^2) \nonumber \\ 
    &- \biggl\{2q^2-(m_1^2+m_2^2) - \frac{(m_1^2-m_2^2)^2}{q^2}\biggr\}B_0^\prime (q^2, m_1^2, m_2^2) 
    \biggr]~. 
\end{align}
For $q^2 = 0$, these functions have simpler forms 
\begin{align}
    B_0' (0, m_1^2, m_2^2) &= \frac{1}{2(m_1^2-m_2^2)^3} \biggl[m_1^4 - m_2^4 + 2 m_1^2 m_2^2 \log \biggl(\frac{m_2^2}{m_1^2}\biggr)\biggr] ~, \\
    B_0' (0, m^2, m^2) &= \frac{1}{6 m^2} ~, \label{eq:b0d1q0mm}\\ 
    H' (0, m_1^2, m_2^2) &=  \frac{1}{18(m_1^2-m_2^2)^3} \biggl[(m_1^2 - m_2^2) \left\{5 (m_1^4+m_2^4) - 22m_1^2m_2^2\right\} \nonumber \\ 
    &- 12 (m_1^2-m_2^2)^2 \left\{m_1^2 \ln \biggl(\frac{m_1^2}{Q^2}\biggr)- m_2^2 \ln \biggl(\frac{m_2^2}{Q^2}\biggr) \right\} \nonumber \\ 
    &- 6 m_1^2 m_2^2 (m_1^2+m_2^2) \ln \biggl(\frac{m_2^2}{m_1^2}\biggr)
    \biggr]~,
    \\
    H' (0, m^2, m^2) &
    = -\frac{1}{3} -\frac{2}{3} \ln \biggl(\frac{m^2}{Q^2} \biggr) ~. \label{eq:h0d1q0mm}
\end{align}

The second derivatives of the functions $B_0 (q^2, m_1^2, m_2^2)$ and $H_0 (q^2, m_1^2, m_2^2)$ with respect to $q^2$ are
\begin{align}
    B_0^{\prime \prime} (q^2, m_1^2, m_2^2) &\equiv
    \frac{d^2}{d(q^2)^2} B_0 (q^2, m_1^2, m_2^2) = \int_0^1 dx \, \frac{x^2(1-x)^2 }{[x m_1^2 + (1-x) m_2^2 - x (1-x) q^2 - i\epsilon]^2} ~, \\
    H^{\prime \prime} (q^2, m_1^2, m_2^2) &\equiv
    \frac{d^2}{d(q^2)^2} H_0 (q^2, m_1^2, m_2^2) \nonumber \\ 
    &= - \frac{1}{3} \biggl[
         \frac{2m_1^2 (m_1^2 - m_2^2)}{(q^2)^3} \ln \biggl(\frac{m_1^2}{Q^2}\biggr)  - \frac{2m_2^2 (m_1^2 - m_2^2)}{(q^2)^3} \ln \biggl(\frac{m_2^2}{Q^2}\biggr)   \nonumber \\ &- \frac{2(m_1^2-m_2^2)^2}{(q^2)^3}
        +\frac{2(m_1^2-m_2^2)^2}{(q^2)^3} B_0(q^2,m_1^2,m_2^2) \nonumber \\ 
        & - 2\biggl\{2+\frac{(m_1^2-m_2^2)^2}{(q^2)^2}\biggr\} B'_0(q^2,m_1^2,m_2^2) \nonumber \\ 
        &- \biggl\{2q^2-(m_1^2+m_2^2) - \frac{(m_1^2-m_2^2)^2}{q^2}\biggr\}B_0^{\prime\prime} (q^2, m_1^2, m_2^2) 
        \biggr]~. 
\end{align}
In the limit of $q^2 = 0$, these functions lead to 
\begin{align}
    B_0^{\prime \prime} (0, m_1^2, m_2^2) &= \frac{1}{3 (m_1^2 - m_2^2)^5} \biggl[(m_1^2 - m_2^2) (m_1^4 + 10 m_1^2 m_2^2 + m_2^4) \nonumber \\  &+ 6 m_1^2 m_2^2 (m_1^2 + m_2^2) \ln \biggl(\frac{m_2^2}{m_1^2}\biggr)\biggr] ~, \\ 
    B_0^{\prime \prime} (0, m^2, m^2) &= \frac{1}{30 m^4} ~, \\ 
    H^{\prime \prime } (0, m_1^2, m_2^2) &= \frac{1}{9(m_1^2-m_2^2)^5} \biggl[6m_1^2m_2^2 (m_1^4 - 6m_1^2m_2^2 + m_2^4) \ln \biggl(\frac{m_2^2}{m_1^2}\biggr)
    \nonumber \\ 
    &+5(m_1^8-m_2^8) -22m_1^2m_2^2 (m_1^4-m_2^4)
    \biggr]~, \\ 
    H^{\prime \prime } (0, m^2, m^2) &=\frac{1}{5 m^2} ~.
\end{align}


\bibliographystyle{utphysmod}
\bibliography{ref}

\providecommand{\href}[2]{#2}\begingroup\raggedright\begin{thebibliography}{100}

\bibitem{Ellis:1986yg}
J.~R.~Ellis, K.~Enqvist, D.~V.~Nanopoulos, and F.~Zwirner, {\em {Observables in Low-Energy Superstring Models}}, \href{https://dx.doi.org/10.1142/S0217732386000105}{Mod.\  Phys.\  Lett.\  A {\bfseries 1} (1986) 57}.

\bibitem{Barbieri:1987fn}
R.~Barbieri and G.~F.~Giudice, {\em {Upper Bounds on Supersymmetric Particle Masses}}, \href{https://dx.doi.org/10.1016/0550-3213(88)90171-X}{Nucl.\  Phys.\  B {\bfseries 306} (1988) 63--76}.

\bibitem{Kitano:2005wc}
R.~Kitano and Y.~Nomura, {\em {A Solution to the supersymmetric fine-tuning problem within the MSSM}}, \href{https://dx.doi.org/10.1016/j.physletb.2005.10.003}{Phys.\  Lett.\  B {\bfseries 631} (2005) 58--67} {\ttfamily [\href{https://arxiv.org/abs/hep-ph/0509039}{hep-ph/0509039}]}.

\bibitem{Kitano:2006gv}
R.~Kitano and Y.~Nomura, {\em {Supersymmetry, naturalness, and signatures at the LHC}}, \href{https://dx.doi.org/10.1103/PhysRevD.73.095004}{Phys.\  Rev.\  D {\bfseries 73} (2006) 095004} {\ttfamily [\href{https://arxiv.org/abs/hep-ph/0602096}{hep-ph/0602096}]}.

\bibitem{Baer:2012up}
H.~Baer, V.~Barger, P.~Huang, A.~Mustafayev, and X.~Tata, {\em {Radiative natural SUSY with a 125 GeV Higgs boson}}, \href{https://dx.doi.org/10.1103/PhysRevLett.109.161802}{Phys.\  Rev.\  Lett.\  {\bfseries 109} (2012) 161802} {\ttfamily [\href{https://arxiv.org/abs/1207.3343}{arXiv:1207.3343}]}.

\bibitem{Baer:2012cf}
H.~Baer, {\em et al.}, {\em {Radiative natural supersymmetry: Reconciling electroweak fine-tuning and the Higgs boson mass}}, \href{https://dx.doi.org/10.1103/PhysRevD.87.115028}{Phys.\  Rev.\  D {\bfseries 87} (2013) 115028} {\ttfamily [\href{https://arxiv.org/abs/1212.2655}{arXiv:1212.2655}]}.

\bibitem{Gherghetta:1999sw}
T.~Gherghetta, G.~F.~Giudice, and J.~D.~Wells, {\em {Phenomenological consequences of supersymmetry with anomaly induced masses}}, \href{https://dx.doi.org/10.1016/S0550-3213(99)00429-0}{Nucl.\  Phys.\  B {\bfseries 559} (1999) 27--47} {\ttfamily [\href{https://arxiv.org/abs/hep-ph/9904378}{hep-ph/9904378}]}.

\bibitem{Moroi:1999zb}
T.~Moroi and L.~Randall, {\em {Wino cold dark matter from anomaly mediated SUSY breaking}}, \href{https://dx.doi.org/10.1016/S0550-3213(99)00748-8}{Nucl.\  Phys.\  B {\bfseries 570} (2000) 455--472} {\ttfamily [\href{https://arxiv.org/abs/hep-ph/9906527}{hep-ph/9906527}]}.

\bibitem{Fujii:2001xp}
M.~Fujii and K.~Hamaguchi, {\em {Higgsino and wino dark matter from Q ball decay}}, \href{https://dx.doi.org/10.1016/S0370-2693(01)01412-5}{Phys.\  Lett.\  B {\bfseries 525} (2002) 143--149} {\ttfamily [\href{https://arxiv.org/abs/hep-ph/0110072}{hep-ph/0110072}]}.

\bibitem{Gelmini:2006pw}
G.~B.~Gelmini and P.~Gondolo, {\em {Neutralino with the right cold dark matter abundance in (almost) any supersymmetric model}}, \href{https://dx.doi.org/10.1103/PhysRevD.74.023510}{Phys.\  Rev.\  D {\bfseries 74} (2006) 023510} {\ttfamily [\href{https://arxiv.org/abs/hep-ph/0602230}{hep-ph/0602230}]}.

\bibitem{Gelmini:2006pq}
G.~Gelmini, P.~Gondolo, A.~Soldatenko, and C.~E.~Yaguna, {\em {The Effect of a late decaying scalar on the neutralino relic density}}, \href{https://dx.doi.org/10.1103/PhysRevD.74.083514}{Phys.\  Rev.\  D {\bfseries 74} (2006) 083514} {\ttfamily [\href{https://arxiv.org/abs/hep-ph/0605016}{hep-ph/0605016}]}.

\bibitem{Baer:2014eja}
H.~Baer, K.-Y.~Choi, J.~E.~Kim, and L.~Roszkowski, {\em {Dark matter production in the early Universe: beyond the thermal WIMP paradigm}}, \href{https://dx.doi.org/10.1016/j.physrep.2014.10.002}{Phys.\  Rept.\  {\bfseries 555} (2015) 1--60} {\ttfamily [\href{https://arxiv.org/abs/1407.0017}{arXiv:1407.0017}]}.

\bibitem{Han:2019vxi}
C.~Han, {\em {Higgsino Dark Matter in a Non-Standard History of the Universe}}, \href{https://dx.doi.org/10.1016/j.physletb.2019.134997}{Phys.\  Lett.\  B {\bfseries 798} (2019) 134997} {\ttfamily [\href{https://arxiv.org/abs/1907.09235}{arXiv:1907.09235}]}.

\bibitem{Fukuda:2024ddb}
H.~Fukuda, Q.~Li, T.~Moroi, and A.~Niki, {\em {Non-thermal production of Higgsino dark matter by late-decaying scalar fields}}, \href{https://dx.doi.org/10.1007/JHEP06(2025)091}{JHEP {\bfseries 06} (2025) 091} {\ttfamily [\href{https://arxiv.org/abs/2410.15733}{arXiv:2410.15733}]}.

\bibitem{Bae:2013bva}
K.~J.~Bae, H.~Baer, and E.~J.~Chun, {\em {Mainly axion cold dark matter from natural supersymmetry}}, \href{https://dx.doi.org/10.1103/PhysRevD.89.031701}{Phys.\  Rev.\  D {\bfseries 89} (2014) 031701} {\ttfamily [\href{https://arxiv.org/abs/1309.0519}{arXiv:1309.0519}]}.

\bibitem{Bae:2013hma}
K.~J.~Bae, H.~Baer, and E.~J.~Chun, {\em {Mixed axion/neutralino dark matter in the SUSY DFSZ axion model}}, \href{https://dx.doi.org/10.1088/1475-7516/2013/12/028}{JCAP {\bfseries 12} (2013) 028} {\ttfamily [\href{https://arxiv.org/abs/1309.5365}{arXiv:1309.5365}]}.

\bibitem{Bae:2017hlp}
K.~J.~Bae, H.~Baer, and H.~Serce, {\em {Prospects for axion detection in natural SUSY with mixed axion-higgsino dark matter: back to invisible?}}, \href{https://dx.doi.org/10.1088/1475-7516/2017/06/024}{JCAP {\bfseries 06} (2017) 024} {\ttfamily [\href{https://arxiv.org/abs/1705.01134}{arXiv:1705.01134}]}.

\bibitem{Nagata:2014wma}
N.~Nagata and S.~Shirai, {\em {Higgsino Dark Matter in High-Scale Supersymmetry}}, \href{https://dx.doi.org/10.1007/JHEP01(2015)029}{JHEP {\bfseries 01} (2015) 029} {\ttfamily [\href{https://arxiv.org/abs/1410.4549}{arXiv:1410.4549}]}.

\bibitem{Fukuda:2017jmk}
H.~Fukuda, N.~Nagata, H.~Otono, and S.~Shirai, {\em {Higgsino Dark Matter or Not: Role of Disappearing Track Searches at the LHC and Future Colliders}}, \href{https://dx.doi.org/10.1016/j.physletb.2018.03.088}{Phys.\  Lett.\  B {\bfseries 781} (2018) 306--311} {\ttfamily [\href{https://arxiv.org/abs/1703.09675}{arXiv:1703.09675}]}.

\bibitem{Krall:2017xij}
R.~Krall and M.~Reece, {\em {Last Electroweak WIMP Standing: Pseudo-Dirac Higgsino Status and Compact Stars as Future Probes}}, \href{https://dx.doi.org/10.1088/1674-1137/42/4/043105}{Chin.\  Phys.\  C {\bfseries 42} (2018) 043105} {\ttfamily [\href{https://arxiv.org/abs/1705.04843}{arXiv:1705.04843}]}.

\bibitem{Fukuda:2019kbp}
H.~Fukuda, N.~Nagata, H.~Oide, H.~Otono, and S.~Shirai, {\em {Cornering Higgsinos Using Soft Displaced Tracks}}, \href{https://dx.doi.org/10.1103/PhysRevLett.124.101801}{Phys.\  Rev.\  Lett.\  {\bfseries 124} (2020) 101801} {\ttfamily [\href{https://arxiv.org/abs/1910.08065}{arXiv:1910.08065}]}.

\bibitem{Canepa:2020ntc}
A.~Canepa, T.~Han, and X.~Wang, {\em {The Search for Electroweakinos}}, \href{https://dx.doi.org/10.1146/annurev-nucl-031020-121031}{Ann.\  Rev.\  Nucl.\  Part.\  Sci.\  {\bfseries 70} (2020) 425--454} {\ttfamily [\href{https://arxiv.org/abs/2003.05450}{arXiv:2003.05450}]}.

\bibitem{Martin:2024pxx}
S.~P.~Martin, {\em {Implications of purity constraints on light Higgsinos}}, \href{https://dx.doi.org/10.1103/PhysRevD.109.095045}{Phys.\  Rev.\  D {\bfseries 109} (2024) 095045} {\ttfamily [\href{https://arxiv.org/abs/2403.19598}{arXiv:2403.19598}]}.

\bibitem{Martin:2024ytt}
S.~P.~Martin, {\em {Curtain lowers on directly detectable higgsino dark matter}}, \href{https://dx.doi.org/10.1103/PhysRevD.111.075004}{Phys.\  Rev.\  D {\bfseries 111} (2025) 075004} {\ttfamily [\href{https://arxiv.org/abs/2412.08958}{arXiv:2412.08958}]}.

\bibitem{ATLAS:2019lng}
{\bfseries ATLAS} Collaboration, {\em {Searches for electroweak production of supersymmetric particles with compressed mass spectra in $\sqrt{s}=$ 13 TeV $pp$ collisions with the ATLAS detector}}, \href{https://dx.doi.org/10.1103/PhysRevD.101.052005}{Phys.\  Rev.\  D {\bfseries 101} (2020) 052005} {\ttfamily [\href{https://arxiv.org/abs/1911.12606}{arXiv:1911.12606}]}.

\bibitem{ATLAS:2022rme}
{\bfseries ATLAS} Collaboration, {\em {Search for long-lived charginos based on a disappearing-track signature using 136 fb$^{-1}$ of pp collisions at $\sqrt{s}$~=~13~TeV with the ATLAS detector}}, \href{https://dx.doi.org/10.1140/epjc/s10052-022-10489-5}{Eur.\  Phys.\  J.\  C {\bfseries 82} (2022) 606} {\ttfamily [\href{https://arxiv.org/abs/2201.02472}{arXiv:2201.02472}]}.

\bibitem{CMS:2023mny}
{\bfseries CMS} Collaboration, {\em {Search for supersymmetry in final states with disappearing tracks in proton-proton collisions at s=13\,\,TeV}}, \href{https://dx.doi.org/10.1103/PhysRevD.109.072007}{Phys.\  Rev.\  D {\bfseries 109} (2024) 072007} {\ttfamily [\href{https://arxiv.org/abs/2309.16823}{arXiv:2309.16823}]}.

\bibitem{ATLAS:2024umc}
{\bfseries ATLAS} Collaboration, {\em {Search for Nearly Mass-Degenerate Higgsinos Using Low-Momentum Mildly Displaced Tracks in pp Collisions at s=13\,\,TeV with the ATLAS Detector}}, \href{https://dx.doi.org/10.1103/PhysRevLett.132.221801}{Phys.\  Rev.\  Lett.\  {\bfseries 132} (2024) 221801} {\ttfamily [\href{https://arxiv.org/abs/2401.14046}{arXiv:2401.14046}]}.

\bibitem{PardodeVera:2020zlr}
M.~T. N.~n.~Pardo~de Vera, M.~Berggren, and J.~List in {\em {International Workshop on Future Linear Colliders}}.
\newblock 2020.
\newblock {\ttfamily \href{https://arxiv.org/abs/2002.01239}{arXiv:2002.01239}}.

\bibitem{ILC:2013jhg}
{\bfseries ILC} Collaboration, {\em {The International Linear Collider Technical Design Report - Volume 2: Physics}}, {\ttfamily \href{https://arxiv.org/abs/1306.6352}{arXiv:1306.6352}} (2013).

\bibitem{ILCInternationalDevelopmentTeam:2022izu}
{\bfseries ILC International Development Team} Collaboration, {\em {The International Linear Collider: Report to Snowmass 2021}}, {\ttfamily \href{https://arxiv.org/abs/2203.07622}{arXiv:2203.07622}} (2022).

\bibitem{Berggren:2013vfa}
M.~Berggren, {\em et al.}, {\em {Tackling light higgsinos at the ILC}}, \href{https://dx.doi.org/10.1140/epjc/s10052-013-2660-y}{Eur.\  Phys.\  J.\  C {\bfseries 73} (2013) 2660} {\ttfamily [\href{https://arxiv.org/abs/1307.3566}{arXiv:1307.3566}]}.

\bibitem{Baer:2014yta}
H.~Baer, V.~Barger, D.~Mickelson, A.~Mustafayev, and X.~Tata, {\em {Physics at a Higgsino Factory}}, \href{https://dx.doi.org/10.1007/JHEP06(2014)172}{JHEP {\bfseries 06} (2014) 172} {\ttfamily [\href{https://arxiv.org/abs/1404.7510}{arXiv:1404.7510}]}.

\bibitem{Moortgat-Pick:2015lbx}
A.~Arbey {\em et~al.}, {\em {Physics at the e+ e- Linear Collider}}, \href{https://dx.doi.org/10.1140/epjc/s10052-015-3511-9}{Eur.\  Phys.\  J.\  C {\bfseries 75} (2015) 371} {\ttfamily [\href{https://arxiv.org/abs/1504.01726}{arXiv:1504.01726}]}.

\bibitem{Bae:2016dxc}
K.~J.~Bae, H.~Baer, N.~Nagata, and H.~Serce, {\em {Precision gaugino mass measurements as a probe of large trilinear soft terms at the ILC}}, \href{https://dx.doi.org/10.1103/PhysRevD.94.035015}{Phys.\  Rev.\  D {\bfseries 94} (2016) 035015} {\ttfamily [\href{https://arxiv.org/abs/1606.02361}{arXiv:1606.02361}]}.

\bibitem{Baer:2019gvu}
H.~Baer, {\em et al.}, {\em {ILC as a natural SUSY discovery machine and precision microscope: From light Higgsinos to tests of unification}}, \href{https://dx.doi.org/10.1103/PhysRevD.101.095026}{Phys.\  Rev.\  D {\bfseries 101} (2020) 095026} {\ttfamily [\href{https://arxiv.org/abs/1912.06643}{arXiv:1912.06643}]}.

\bibitem{Harigaya:2015yaa}
K.~Harigaya, K.~Ichikawa, A.~Kundu, S.~Matsumoto, and S.~Shirai, {\em {Indirect Probe of Electroweak-Interacting Particles at Future Lepton Colliders}}, \href{https://dx.doi.org/10.1007/JHEP09(2015)105}{JHEP {\bfseries 09} (2015) 105} {\ttfamily [\href{https://arxiv.org/abs/1504.03402}{arXiv:1504.03402}]}.

\bibitem{Liu:2017msv}
N.~Liu and L.~Wu, {\em {An indirect probe of the higgsino world at the CEPC}}, \href{https://dx.doi.org/10.1140/epjc/s10052-017-5443-z}{Eur.\  Phys.\  J.\  C {\bfseries 77} (2017) 868} {\ttfamily [\href{https://arxiv.org/abs/1705.02534}{arXiv:1705.02534}]}.

\bibitem{DiLuzio:2018jwd}
L.~Di~Luzio, R.~Gr\"ober, and G.~Panico, {\em {Probing new electroweak states via precision measurements at the LHC and future colliders}}, \href{https://dx.doi.org/10.1007/JHEP01(2019)011}{JHEP {\bfseries 01} (2019) 011} {\ttfamily [\href{https://arxiv.org/abs/1810.10993}{arXiv:1810.10993}]}.

\bibitem{Gao:2021jip}
L.-Q.~Gao, X.-J.~Bi, J.-W.~Wang, Q.-F.~Xiang, and P.-F.~Yin, {\em {Exploring fermionic multiplet dark matter through precision measurements at the CEPC *}}, \href{https://dx.doi.org/10.1088/1674-1137/ac7547}{Chin.\  Phys.\  C {\bfseries 46} (2022) 093112} {\ttfamily [\href{https://arxiv.org/abs/2112.02519}{arXiv:2112.02519}]}.

\bibitem{FCC:2018evy}
{\bfseries FCC} Collaboration, {\em {FCC-ee: The Lepton Collider}: {Future Circular Collider Conceptual Design Report Volume 2}}, \href{https://dx.doi.org/10.1140/epjst/e2019-900045-4}{Eur.\  Phys.\  J.\  ST {\bfseries 228} (2019) 261--623}.

\bibitem{Bernardi:2022hny}
G.~Bernardi {\em et~al.}, {\em {The Future Circular Collider: a Summary for the US 2021 Snowmass Process}}, {\ttfamily \href{https://arxiv.org/abs/2203.06520}{arXiv:2203.06520}} (2022).

\bibitem{CEPCStudyGroup:2018ghi}
{\bfseries CEPC Study Group} Collaboration, {\em {CEPC Conceptual Design Report: Volume 2 - Physics \& Detector}}, {\ttfamily \href{https://arxiv.org/abs/1811.10545}{arXiv:1811.10545}} (2018).

\bibitem{CEPCPhysicsStudyGroup:2022uwl}
{\bfseries CEPC Physics Study Group} Collaboration in {\em {Snowmass 2021}}.
\newblock 2022.
\newblock {\ttfamily \href{https://arxiv.org/abs/2205.08553}{arXiv:2205.08553}}.

\bibitem{Maura:2024zxz}
V.~Maura, B.~A.~Stefanek, and T.~You, {\em {Accuracy complements energy: electroweak precision tests at Tera-Z}}, {\ttfamily \href{https://arxiv.org/abs/2412.14241}{arXiv:2412.14241}} (2024).

\bibitem{Fan:2014vta}
J.~Fan, M.~Reece, and L.-T.~Wang, {\em {Possible Futures of Electroweak Precision: ILC, FCC-ee, and CEPC}}, \href{https://dx.doi.org/10.1007/JHEP09(2015)196}{JHEP {\bfseries 09} (2015) 196} {\ttfamily [\href{https://arxiv.org/abs/1411.1054}{arXiv:1411.1054}]}.

\bibitem{Fan:2014axa}
J.~Fan, M.~Reece, and L.-T.~Wang, {\em {Precision Natural SUSY at CEPC, FCC-ee, and ILC}}, \href{https://dx.doi.org/10.1007/JHEP08(2015)152}{JHEP {\bfseries 08} (2015) 152} {\ttfamily [\href{https://arxiv.org/abs/1412.3107}{arXiv:1412.3107}]}.

\bibitem{Cai:2016sjz}
C.~Cai, Z.-H.~Yu, and H.-H.~Zhang, {\em {CEPC Precision of Electroweak Oblique Parameters and Weakly Interacting Dark Matter: the Fermionic Case}}, \href{https://dx.doi.org/10.1016/j.nuclphysb.2017.05.015}{Nucl.\  Phys.\  B {\bfseries 921} (2017) 181--210} {\ttfamily [\href{https://arxiv.org/abs/1611.02186}{arXiv:1611.02186}]}.

\bibitem{Cai:2017wdu}
C.~Cai, Z.-H.~Yu, and H.-H.~Zhang, {\em {CEPC Precision of Electroweak Oblique Parameters and Weakly Interacting Dark Matter: the Scalar Case}}, \href{https://dx.doi.org/10.1016/j.nuclphysb.2017.09.007}{Nucl.\  Phys.\  B {\bfseries 924} (2017) 128--152} {\ttfamily [\href{https://arxiv.org/abs/1705.07921}{arXiv:1705.07921}]}.

\bibitem{Dawson:2022bxd}
S.~Dawson and P.~P.~Giardino, {\em {Flavorful electroweak precision observables in the Standard Model effective field theory}}, \href{https://dx.doi.org/10.1103/PhysRevD.105.073006}{Phys.\  Rev.\  D {\bfseries 105} (2022) 073006} {\ttfamily [\href{https://arxiv.org/abs/2201.09887}{arXiv:2201.09887}]}.

\bibitem{Bottaro:2022one}
S.~Bottaro, {\em et al.}, {\em {The last complex WIMPs standing}}, \href{https://dx.doi.org/10.1140/epjc/s10052-022-10918-5}{Eur.\  Phys.\  J.\  C {\bfseries 82} (2022) 992} {\ttfamily [\href{https://arxiv.org/abs/2205.04486}{arXiv:2205.04486}]}.

\bibitem{deBlas:2022ofj}
J.~de~Blas, {\em et al.} in {\em {Snowmass 2021}}.
\newblock 2022.
\newblock {\ttfamily \href{https://arxiv.org/abs/2206.08326}{arXiv:2206.08326}}.

\bibitem{Allwicher:2023aql}
L.~Allwicher, G.~Isidori, J.~M.~Lizana, N.~Selimovic, and B.~A.~Stefanek, {\em {Third-family quark-lepton Unification and electroweak precision tests}}, \href{https://dx.doi.org/10.1007/JHEP05(2023)179}{JHEP {\bfseries 05} (2023) 179} {\ttfamily [\href{https://arxiv.org/abs/2302.11584}{arXiv:2302.11584}]}.

\bibitem{Bellafronte:2023amz}
L.~Bellafronte, S.~Dawson, and P.~P.~Giardino, {\em {The importance of flavor in SMEFT Electroweak Precision Fits}}, \href{https://dx.doi.org/10.1007/JHEP05(2023)208}{JHEP {\bfseries 05} (2023) 208} {\ttfamily [\href{https://arxiv.org/abs/2304.00029}{arXiv:2304.00029}]}.

\bibitem{Allwicher:2023shc}
L.~Allwicher, C.~Cornella, G.~Isidori, and B.~A.~Stefanek, {\em {New physics in the third generation. A comprehensive SMEFT analysis and future prospects}}, \href{https://dx.doi.org/10.1007/JHEP03(2024)049}{JHEP {\bfseries 03} (2024) 049} {\ttfamily [\href{https://arxiv.org/abs/2311.00020}{arXiv:2311.00020}]}.

\bibitem{Celada:2024mcf}
E.~Celada, {\em et al.}, {\em {Mapping the SMEFT at high-energy colliders: from LEP and the (HL-)LHC to the FCC-ee}}, \href{https://dx.doi.org/10.1007/JHEP09(2024)091}{JHEP {\bfseries 09} (2024) 091} {\ttfamily [\href{https://arxiv.org/abs/2404.12809}{arXiv:2404.12809}]}.

\bibitem{Stefanek:2024kds}
B.~A.~Stefanek, {\em {Non-universal probes of composite Higgs models: new bounds and prospects for FCC-ee}}, \href{https://dx.doi.org/10.1007/JHEP09(2024)103}{JHEP {\bfseries 09} (2024) 103} {\ttfamily [\href{https://arxiv.org/abs/2407.09593}{arXiv:2407.09593}]}.

\bibitem{Knapen:2024bxw}
S.~Knapen, K.~Langhoff, and Z.~Ligeti, {\em {Imprints of supersymmetry at a future Z factory}}, \href{https://dx.doi.org/10.1103/fw5y-svmh}{Phys.\  Rev.\  D {\bfseries 111} (2025) 115007} {\ttfamily [\href{https://arxiv.org/abs/2407.13815}{arXiv:2407.13815}]}.

\bibitem{Allwicher:2024sso}
L.~Allwicher, M.~McCullough, and S.~Renner, {\em {New physics at Tera-Z: precision renormalised}}, \href{https://dx.doi.org/10.1007/JHEP02(2025)164}{JHEP {\bfseries 02} (2025) 164} {\ttfamily [\href{https://arxiv.org/abs/2408.03992}{arXiv:2408.03992}]}.

\bibitem{Crawford:2024nun}
G.~Crawford and D.~Sutherland, {\em {Scalars with non-decoupling phenomenology at future colliders}}, \href{https://dx.doi.org/10.1007/JHEP04(2025)197}{JHEP {\bfseries 04} (2025) 197} {\ttfamily [\href{https://arxiv.org/abs/2409.18177}{arXiv:2409.18177}]}.

\bibitem{Erdelyi:2024sls}
B.~A.~Erdelyi, R.~Gr\"ober, and N.~Selimovic, {\em {How large can the light quark Yukawa couplings be?}}, \href{https://dx.doi.org/10.1007/JHEP05(2025)189}{JHEP {\bfseries 05} (2025) 189} {\ttfamily [\href{https://arxiv.org/abs/2410.08272}{arXiv:2410.08272}]}.

\bibitem{Greljo:2024ytg}
A.~Greljo, H.~Tiblom, and A.~Valenti, {\em {New Physics Through Flavor Tagging at FCC-ee}}, {\ttfamily \href{https://arxiv.org/abs/2411.02485}{arXiv:2411.02485}} (2024).

\bibitem{Gargalionis:2024jaw}
J.~Gargalionis, J.~Quevillon, P.~N.~H.~Vuong, and T.~You, {\em {Linear Standard Model extensions in the SMEFT at one loop and Tera-Z}}, {\ttfamily \href{https://arxiv.org/abs/2412.01759}{arXiv:2412.01759}} (2024).

\bibitem{Davighi:2024syj}
J.~Davighi, {\em {In Search of an Invisible $Z^\prime$}}, {\ttfamily \href{https://arxiv.org/abs/2412.07694}{arXiv:2412.07694}} (2024).

\bibitem{Erdelyi:2025axy}
B.~A.~Erdelyi, R.~Gr\"ober, and N.~Selimovic, {\em {Probing new physics with the electron Yukawa coupling}}, \href{https://dx.doi.org/10.1007/JHEP05(2025)135}{JHEP {\bfseries 05} (2025) 135} {\ttfamily [\href{https://arxiv.org/abs/2501.07628}{arXiv:2501.07628}]}.

\bibitem{Peskin:1991sw}
M.~E.~Peskin and T.~Takeuchi, {\em {Estimation of oblique electroweak corrections}}, \href{https://dx.doi.org/10.1103/PhysRevD.46.381}{Phys.\  Rev.\  D {\bfseries 46} (1992) 381--409}.

\bibitem{Maksymyk:1993zm}
I.~Maksymyk, C.~P.~Burgess, and D.~London, {\em {Beyond S, T and U}}, \href{https://dx.doi.org/10.1103/PhysRevD.50.529}{Phys.\  Rev.\  D {\bfseries 50} (1994) 529--535} {\ttfamily [\href{https://arxiv.org/abs/hep-ph/9306267}{hep-ph/9306267}]}.

\bibitem{Barbieri:2004qk}
R.~Barbieri, A.~Pomarol, R.~Rattazzi, and A.~Strumia, {\em {Electroweak symmetry breaking after LEP-1 and LEP-2}}, \href{https://dx.doi.org/10.1016/j.nuclphysb.2004.10.014}{Nucl.\  Phys.\  B {\bfseries 703} (2004) 127--146} {\ttfamily [\href{https://arxiv.org/abs/hep-ph/0405040}{hep-ph/0405040}]}.

\bibitem{Martin:2004id}
S.~P.~Martin, K.~Tobe, and J.~D.~Wells, {\em {Virtual effects of light gauginos and higgsinos: A Precision electroweak analysis of split supersymmetry}}, \href{https://dx.doi.org/10.1103/PhysRevD.71.073014}{Phys.\  Rev.\  D {\bfseries 71} (2005) 073014} {\ttfamily [\href{https://arxiv.org/abs/hep-ph/0412424}{hep-ph/0412424}]}.

\bibitem{Pierce:1996zz}
D.~M.~Pierce, J.~A.~Bagger, K.~T.~Matchev, and R.-j.~Zhang, {\em {Precision corrections in the minimal supersymmetric standard model}}, \href{https://dx.doi.org/10.1016/S0550-3213(96)00683-9}{Nucl.\  Phys.\  B {\bfseries 491} (1997) 3--67} {\ttfamily [\href{https://arxiv.org/abs/hep-ph/9606211}{hep-ph/9606211}]}.

\bibitem{Wells:2015uba}
J.~D.~Wells and Z.~Zhang, {\em {Effective theories of universal theories}}, \href{https://dx.doi.org/10.1007/JHEP01(2016)123}{JHEP {\bfseries 01} (2016) 123} {\ttfamily [\href{https://arxiv.org/abs/1510.08462}{arXiv:1510.08462}]}.

\bibitem{Marandella:2005wc}
G.~Marandella, C.~Schappacher, and A.~Strumia, {\em {Supersymmetry and precision data after LEP2}}, \href{https://dx.doi.org/10.1016/j.nuclphysb.2005.03.001}{Nucl.\  Phys.\  B {\bfseries 715} (2005) 173--189} {\ttfamily [\href{https://arxiv.org/abs/hep-ph/0502095}{hep-ph/0502095}]}.

\bibitem{Cheung:2012qy}
C.~Cheung, L.~J.~Hall, D.~Pinner, and J.~T.~Ruderman, {\em {Prospects and Blind Spots for Neutralino Dark Matter}}, \href{https://dx.doi.org/10.1007/JHEP05(2013)100}{JHEP {\bfseries 05} (2013) 100} {\ttfamily [\href{https://arxiv.org/abs/1211.4873}{arXiv:1211.4873}]}.

\bibitem{ParticleDataGroup:2024cfk}
{\bfseries Particle Data Group} Collaboration, {\em {Review of particle physics}}, \href{https://dx.doi.org/10.1103/PhysRevD.110.030001}{Phys.\  Rev.\  D {\bfseries 110} (2024) 030001}.

\bibitem{Grifols:1984xs}
J.~A.~Grifols and J.~Sola, {\em {One Loop Renormalization of the Electroweak Parameters in $N=1$ Supersymmetry}}, \href{https://dx.doi.org/10.1016/0550-3213(85)90519-X}{Nucl.\  Phys.\  B {\bfseries 253} (1985) 47}.

\bibitem{Barbieri:1989dc}
R.~Barbieri, M.~Frigeni, F.~Giuliani, and H.~E.~Haber, {\em {Precision Measurements in Electroweak Physics and Supersymmetry}}, \href{https://dx.doi.org/10.1016/0550-3213(90)90181-C}{Nucl.\  Phys.\  B {\bfseries 341} (1990) 309--321}.

\bibitem{Drees:1990dx}
M.~Drees and K.~Hagiwara, {\em {Supersymmetric Contribution to the Electroweak $\rho$ Parameter}}, \href{https://dx.doi.org/10.1103/PhysRevD.42.1709}{Phys.\  Rev.\  D {\bfseries 42} (1990) 1709--1725}.

\bibitem{Drees:1991zk}
M.~Drees, K.~Hagiwara, and A.~Yamada, {\em {Process independent radiative corrections in the minimal supersymmetric standard model}}, \href{https://dx.doi.org/10.1103/PhysRevD.45.1725}{Phys.\  Rev.\  D {\bfseries 45} (1992) 1725--1743}.

\bibitem{Barbieri:1991qp}
R.~Barbieri, M.~Frigeni, and F.~Caravaglios, {\em {Supersymmetry signals in electroweak precision tests at LEP}}, \href{https://dx.doi.org/10.1016/0370-2693(92)91860-C}{Phys.\  Lett.\  B {\bfseries 279} (1992) 169--176}.

\bibitem{Chankowski:1993eu}
P.~H.~Chankowski, {\em et al.}, {\em {Delta R in the MSSM}}, \href{https://dx.doi.org/10.1016/0550-3213(94)90539-8}{Nucl.\  Phys.\  B {\bfseries 417} (1994) 101--129}.

\bibitem{Cho:1999km}
G.-C.~Cho and K.~Hagiwara, {\em {Supersymmetry versus precision experiments revisited}}, \href{https://dx.doi.org/10.1016/S0550-3213(00)00027-4}{Nucl.\  Phys.\  B {\bfseries 574} (2000) 623--674} {\ttfamily [\href{https://arxiv.org/abs/hep-ph/9912260}{hep-ph/9912260}]}.

\bibitem{Heinemeyer:2004gx}
S.~Heinemeyer, W.~Hollik, and G.~Weiglein, {\em {Electroweak precision observables in the minimal supersymmetric standard model}}, \href{https://dx.doi.org/10.1016/j.physrep.2005.12.002}{Phys.\  Rept.\  {\bfseries 425} (2006) 265--368} {\ttfamily [\href{https://arxiv.org/abs/hep-ph/0412214}{hep-ph/0412214}]}.

\bibitem{Heinemeyer:2006px}
S.~Heinemeyer, W.~Hollik, D.~Stockinger, A.~M.~Weber, and G.~Weiglein, {\em {Precise prediction for M(W) in the MSSM}}, \href{https://dx.doi.org/10.1088/1126-6708/2006/08/052}{JHEP {\bfseries 08} (2006) 052} {\ttfamily [\href{https://arxiv.org/abs/hep-ph/0604147}{hep-ph/0604147}]}.

\bibitem{Heinemeyer:2007bw}
S.~Heinemeyer, W.~Hollik, A.~M.~Weber, and G.~Weiglein, {\em {$Z$ Pole Observables in the MSSM}}, \href{https://dx.doi.org/10.1088/1126-6708/2008/04/039}{JHEP {\bfseries 04} (2008) 039} {\ttfamily [\href{https://arxiv.org/abs/0710.2972}{arXiv:0710.2972}]}.

\bibitem{Altarelli:1990zd}
G.~Altarelli and R.~Barbieri, {\em {Vacuum polarization effects of new physics on electroweak processes}}, \href{https://dx.doi.org/10.1016/0370-2693(91)91378-9}{Phys.\  Lett.\  B {\bfseries 253} (1991) 161--167}.

\bibitem{Altarelli:1991fk}
G.~Altarelli, R.~Barbieri, and S.~Jadach, {\em {Toward a model independent analysis of electroweak data}}, \href{https://dx.doi.org/10.1016/0550-3213(92)90376-M}{Nucl.\  Phys.\  B {\bfseries 369} (1992) 3--32}. [Erratum: Nucl.Phys.B 376, 444 (1992)].

\bibitem{ATLAS:2018qzr}
{\bfseries ATLAS} Collaboration, {\em {Prospects for the measurement of the W-boson mass at the HL- and HE-LHC}},  (2018).

\bibitem{ATLAS:2018qvs}
{\bfseries ATLAS} Collaboration, {\em {Prospect for a measurement of the Weak Mixing Angle in $pp \rightarrow Z/\gamma^* \rightarrow e^+e^-$ events with the ATLAS detector at the High Luminosity Large Hadron Collider}},  (2018).

\bibitem{CMS:2017vxj}
{\bfseries CMS} Collaboration, {\em {A proposal for the measurement of the weak mixing angle at the HL-LHC}},  (2017).

\bibitem{Mizuno:2022xuk}
T.~Mizuno, K.~Fujii, and J.~Tian in {\em {Snowmass 2021}}.
\newblock 2022.
\newblock {\ttfamily \href{https://arxiv.org/abs/2203.07944}{arXiv:2203.07944}}.

\bibitem{AbdulKhalek:2018rok}
R.~Abdul~Khalek, S.~Bailey, J.~Gao, L.~Harland-Lang, and J.~Rojo, {\em {Towards Ultimate Parton Distributions at the High-Luminosity LHC}}, \href{https://dx.doi.org/10.1140/epjc/s10052-018-6448-y}{Eur.\  Phys.\  J.\  C {\bfseries 78} (2018) 962} {\ttfamily [\href{https://arxiv.org/abs/1810.03639}{arXiv:1810.03639}]}.

\bibitem{Wilson:2016hne}
G.~W.~Wilson in {\em {International Workshop on Future Linear Colliders}}.
\newblock 2016.
\newblock {\ttfamily \href{https://arxiv.org/abs/1603.06016}{arXiv:1603.06016}}.

\bibitem{Yokoya:2019rhx}
K.~Yokoya, K.~Kubo, and T.~Okugi, {\em {Operation of ILC250 at the Z-pole}}, {\ttfamily \href{https://arxiv.org/abs/1908.08212}{arXiv:1908.08212}} (2019).

\bibitem{Wells:2003tf}
J.~D.~Wells in {\em {11th International Conference on Supersymmetry and the Unification of Fundamental Interactions}}.
\newblock 2003.
\newblock {\ttfamily \href{https://arxiv.org/abs/hep-ph/0306127}{hep-ph/0306127}}.

\bibitem{Wells:2004di}
J.~D.~Wells, {\em {PeV-scale supersymmetry}}, \href{https://dx.doi.org/10.1103/PhysRevD.71.015013}{Phys.\  Rev.\  D {\bfseries 71} (2005) 015013} {\ttfamily [\href{https://arxiv.org/abs/hep-ph/0411041}{hep-ph/0411041}]}.

\bibitem{Arkani-Hamed:2004ymt}
N.~Arkani-Hamed and S.~Dimopoulos, {\em {Supersymmetric unification without low energy supersymmetry and signatures for fine-tuning at the LHC}}, \href{https://dx.doi.org/10.1088/1126-6708/2005/06/073}{JHEP {\bfseries 06} (2005) 073} {\ttfamily [\href{https://arxiv.org/abs/hep-th/0405159}{hep-th/0405159}]}.

\bibitem{Giudice:2004tc}
G.~F.~Giudice and A.~Romanino, {\em {Split supersymmetry}}, \href{https://dx.doi.org/10.1016/j.nuclphysb.2004.08.001}{Nucl.\  Phys.\  B {\bfseries 699} (2004) 65--89} {\ttfamily [\href{https://arxiv.org/abs/hep-ph/0406088}{hep-ph/0406088}]}. [Erratum: Nucl.Phys.B 706, 487--487 (2005)].

\bibitem{Arkani-Hamed:2004zhs}
N.~Arkani-Hamed, S.~Dimopoulos, G.~F.~Giudice, and A.~Romanino, {\em {Aspects of split supersymmetry}}, \href{https://dx.doi.org/10.1016/j.nuclphysb.2004.12.026}{Nucl.\  Phys.\  B {\bfseries 709} (2005) 3--46} {\ttfamily [\href{https://arxiv.org/abs/hep-ph/0409232}{hep-ph/0409232}]}.

\bibitem{Arkani-Hamed:2005zuc}
N.~Arkani-Hamed, S.~Dimopoulos, and S.~Kachru, {\em {Predictive landscapes and new physics at a TeV}}, {\ttfamily \href{https://arxiv.org/abs/hep-th/0501082}{hep-th/0501082}} (2005).

\bibitem{Hall:2011jd}
L.~J.~Hall and Y.~Nomura, {\em {Spread Supersymmetry}}, \href{https://dx.doi.org/10.1007/JHEP01(2012)082}{JHEP {\bfseries 01} (2012) 082} {\ttfamily [\href{https://arxiv.org/abs/1111.4519}{arXiv:1111.4519}]}.

\bibitem{Hall:2012zp}
L.~J.~Hall, Y.~Nomura, and S.~Shirai, {\em {Spread Supersymmetry with Wino LSP: Gluino and Dark Matter Signals}}, \href{https://dx.doi.org/10.1007/JHEP01(2013)036}{JHEP {\bfseries 01} (2013) 036} {\ttfamily [\href{https://arxiv.org/abs/1210.2395}{arXiv:1210.2395}]}.

\bibitem{Ibe:2011aa}
M.~Ibe and T.~T.~Yanagida, {\em {The Lightest Higgs Boson Mass in Pure Gravity Mediation Model}}, \href{https://dx.doi.org/10.1016/j.physletb.2012.02.034}{Phys.\  Lett.\  B {\bfseries 709} (2012) 374--380} {\ttfamily [\href{https://arxiv.org/abs/1112.2462}{arXiv:1112.2462}]}.

\bibitem{Ibe:2012hu}
M.~Ibe, S.~Matsumoto, and T.~T.~Yanagida, {\em {Pure Gravity Mediation with $m_{3/2} = 10$--$100$ TeV}}, \href{https://dx.doi.org/10.1103/PhysRevD.85.095011}{Phys.\  Rev.\  D {\bfseries 85} (2012) 095011} {\ttfamily [\href{https://arxiv.org/abs/1202.2253}{arXiv:1202.2253}]}.

\bibitem{Arvanitaki:2012ps}
A.~Arvanitaki, N.~Craig, S.~Dimopoulos, and G.~Villadoro, {\em {Mini-Split}}, \href{https://dx.doi.org/10.1007/JHEP02(2013)126}{JHEP {\bfseries 02} (2013) 126} {\ttfamily [\href{https://arxiv.org/abs/1210.0555}{arXiv:1210.0555}]}.

\bibitem{Arkani-Hamed:2012fhg}
N.~Arkani-Hamed, A.~Gupta, D.~E.~Kaplan, N.~Weiner, and T.~Zorawski, {\em {Simply Unnatural Supersymmetry}}, {\ttfamily \href{https://arxiv.org/abs/1212.6971}{arXiv:1212.6971}} (2012).

\bibitem{Evans:2013lpa}
J.~L.~Evans, M.~Ibe, K.~A.~Olive, and T.~T.~Yanagida, {\em {Universality in Pure Gravity Mediation}}, \href{https://dx.doi.org/10.1140/epjc/s10052-013-2468-9}{Eur.\  Phys.\  J.\  C {\bfseries 73} (2013) 2468} {\ttfamily [\href{https://arxiv.org/abs/1302.5346}{arXiv:1302.5346}]}.

\bibitem{Evans:2013dza}
J.~L.~Evans, K.~A.~Olive, M.~Ibe, and T.~T.~Yanagida, {\em {Non-Universalities in Pure Gravity Mediation}}, \href{https://dx.doi.org/10.1140/epjc/s10052-013-2611-7}{Eur.\  Phys.\  J.\  C {\bfseries 73} (2013) 2611} {\ttfamily [\href{https://arxiv.org/abs/1305.7461}{arXiv:1305.7461}]}.

\end{thebibliography}\endgroup


\end{document}